\documentclass[aps,prc,preprint,tightenlines,floatfix,groupedaddress,
nofootinbib,showpacs,preprintnumbers,amsmath,amssymb,superscriptaddress]
{revtex4}
\usepackage{graphicx}% Include figure files
\usepackage{dcolumn}%Align table columns on decimal point
\usepackage{mathrsfs}
\usepackage{bm}% bold math
\begin{document}

\title{Incompressibility of neutron-rich matter}
\author{J. Piekarewicz\footnote{\tt e-mail: jpiekarewicz@fsu.edu}}
\affiliation{Department of Physics, Florida State University,
             Tallahassee, FL {\sl 32306}}
\author{M. Centelles\footnote{\tt e-mail: mariocentelles@ub.edu}}
\affiliation{Departament d'Estructura i Constituents de la Mat\`eria
and Institut de Ci\`encies del Cosmos, Facultat de F\'{\i}sica,
Universitat de Barcelona, Diagonal {\sl 647}, {\sl 08028} Barcelona,
Spain} 
\date{\today}

\begin{abstract}
The saturation properties of neutron-rich matter are investigated in a
relativistic mean-field formalism using two accurately calibrated
models: NL3 and FSUGold. The saturation properties---{\sl density,
binding energy per nucleon, and incompressibility coefficient}---are
calculated as a function of the neutron-proton asymmetry
$\alpha\!\equiv\!(N\!-\!Z)/A$ to all orders in $\alpha$. Good
agreement (at the 10\% level or better) is found between these
numerical calculations and analytic expansions that are given in terms
of a handful of bulk parameters determined at saturation density.
Using insights developed from the analytic approach and a general
expression for the incompressibility coefficient of infinite
neutron-rich matter, {\sl i.e.,}
$K_{0}(\alpha)=K_{0}+K_{\tau}\alpha^{2}+\ldots$, we construct a Hybrid
model with values for $K_{0}$ and $K_{\tau}$ as suggested by recent
experimental findings. Whereas the Hybrid model provides a better
description of the measured distribution of isoscalar monopole
strength in the Sn-isotopes relative to both NL3 and FSUGold, it
significantly underestimates the distribution of strength in
${}^{208}$Pb. Thus, we conclude that the incompressibility coefficient
of neutron-rich matter remains an important open problem.
 \end{abstract}
\maketitle 

\section{Introduction}
\label{Introduction}

The incompressibility of neutron-rich matter remains at the forefront
of both experimental and theoretical investigations due to of its
fundamental role in constraining the equation of state (EOS) of cold
dense matter. The incompressibility coefficient characterizes the
small-density fluctuations of infinite nuclear matter around its
equilibrium point, thereby providing the first glimpse into the
\textsl{``stiffness"} of the equation of state.

It is widely accepted that the {\sl Giant Monopole Resonance} (GMR)
--- the quintessential compressional mode --- provides the cleanest,
most direct route to the nuclear incompressibility. In a procedure
first proposed by Blaizot and collaborators~\cite{Blaizot:1976,
Blaizot:1995}, finite nuclei GMR energies as well as the
nuclear-matter incompressibility should both be computed within the
same theoretical framework. This procedure avoids altogether the
reliance on macroscopic ({\sl ``liquid-drop-like"}) approaches that
have proven unreliable for the extraction of the incompressibility
coefficient of {\sl infinite nuclear matter} from finite nuclei GMR
energies~\cite{Pearson:1991, Shlomo:1993zz}. In most theoretical
approaches, an accurately calibrated model is obtained by fitting the
model parameters to a set of selected ground-state properties of
finite nuclei. In some recent instances, excited states --- computed
as the consistent linear response of the mean-field ground state ---
have also been incorporated into the fit~\cite{Todd-Rutel:2005fa}.  If
such accurately calibrated models are able to reproduce the
experimental distribution of monopole strength (or at least some of
its moments) then the value of the incompressibility coefficient
predicted by the model is regarded as reliable. Following this
procedure it has been established that the incompressibility
coefficient of \textsl{symmetric} nuclear matter falls within the
following relatively narrow range:
$K_{0}\!=\!240\!\pm\!10$~MeV~\cite{Garg:2006vc,Li:2007bp}.

Because of the collective nature of the GMR, a strong coherent peak
develops only in the case of relatively heavy nuclei. Indeed, the
monopole strength in ``light" nuclei, such as ${}^{40}$Ca, is strongly
fragmented. As (stable) heavy nuclei are characterized by a
significant neutron excess, experimental studies of the GMR probe the
incompressibility of {\sl neutron-rich matter} rather than that of
symmetric matter. As such, GMR energies on a variety of nuclei having
different neutron-proton asymmetries [$\alpha\!\equiv\!(N\!-\!Z)/A$],
such as ${}^{90}$Zr, ${}^{116}$Sn, ${}^{144}$Sm, and ${}^{208}$Pb,
provide \textsl{simultaneous} constraints on the incompressibility of
symmetric nuclear matter ($K_{0}$) as well as on its leading $\alpha$
correction, a quantity that will be denoted by $K_{\tau}$. 
That is, the incompressibility of \textsl{infinite neutron-rich
matter} may be parametrized to leading order in the neutron-proton
asymmetry as $K_{0}(\alpha)=K_{0}+K_{\tau}\alpha^{2}$. It is therefore
natural to assume that previous lessons learned in the case of $K_{0}$ 
will remain relevant for $K_{\tau}$. First and foremost, $K_{\tau}$
should not be inferred from an extrapolation to the
$A\!\rightarrow\infty$ limit from laboratory experiments on finite
nuclei. Rather, one should continue to follow the procedure advocated
by Blaizot and demand that the values of both $K_{0}$ and $K_{\tau}$
be those predicted by a consistent theoretical model that successfully
reproduces the experimental GMR energies of a variety of nuclei.  We
note that in the present contribution both $K_{0}$ and $K_{\tau}$ refer 
to bulk properties of the \textsl{infinite} system.

It is therefore the aim of the present manuscript to: (a) use
accurately calibrated relativistic mean-field models to extract the
saturation properties of \textsl{infinite} neutron-rich matter; (b)
compute GMR energies for a variety of nuclei using the consistent
isoscalar-monopole response of the mean-field ground state; and (c)
confront these theoretical results against experimental GMR energies
--- especially the new data on Tin~\cite{Garg:2006vc,Li:2007bp}. As 
a byproduct of this procedure, analytic approaches to the saturation
properties of infinite neutron-rich matter based on a few bulk 
parameters calculated at saturation density will be validated against 
exact numerical results.

The manuscript has been organized as follows. In Sec.~\ref{Formalism}
a brief review of the relativistic mean-field formalism will be
provided.  Particular emphasis will be placed on developing a thorough
description of the properties of infinite neutron-rich matter and on
the various bulk parameters that define its behavior around
nuclear-matter saturation density. In Sec.~\ref{Results} results will
be presented for the evolution of the saturation point with
neutron-proton asymmetry using both exact (numerical) and approximate
(analytic) approaches. We finish this section by revisiting the topic
of {\sl why is Tin so soft?}  \cite{Garg:2006vc,Li:2007bp,
Piekarewicz:2007us,Sagawa:2007sp,Avdeenkov:2008bi}. Our summary and
conclusions will follow in Sec.~\ref{Conclusions}.

\section{Formalism}
\label{Formalism}

The starting point for the calculation of various ground-state properties
is an interacting Lagrangian density of the following form:
\begin{widetext}
\begin{align}
{\mathscr L}_{\rm int} & =
\bar\psi \left[g_{\rm s}\phi   \!-\!
         \left(g_{\rm v}V_\mu  \!+\!
    \frac{g_{\rho}}{2}\mbox{\boldmath$\tau$}\cdot{\bf b}_{\mu}
                               \!+\!
    \frac{e}{2}(1\!+\!\tau_{3})A_{\mu}\right)\gamma^{\mu}
         \right]\psi \nonumber \\
                   & -
    \frac{\kappa}{3!} (g_{\rm s}\phi)^3 \!-\!
    \frac{\lambda}{4!}(g_{\rm s}\phi)^4 \!+\!
    \frac{\zeta}{4!}
    \Big(g_{\rm v}^2 V_{\mu}V^\mu\Big)^2 \!+\!
    \Lambda_{\rm v}
    \Big(g_{\rho}^{2}\,{\bf b}_{\mu}\cdot{\bf b}^{\mu}\Big)
    \Big(g_{\rm v}^2V_{\mu}V^\mu\Big) \;.
 \label{Lagrangian}
\end{align}
\end{widetext}
The Lagrangian density includes an isodoublet nucleon field ($\psi$)
interacting via the exchange of two isoscalar mesons, a scalar
($\phi$) and a vector ($V^{\mu}$), one isovector meson ($b^{\mu}$),
and the photon ($A^{\mu}$)~\cite{Serot:1984ey,Serot:1997xg}.  In
addition to meson-nucleon interactions, the Lagrangian density is
supplemented by nonlinear meson interactions with coupling constants
denoted by $\kappa$, $\lambda$, $\zeta$, and $\Lambda_{\rm v}$ that
are responsible for a softening of the nuclear-matter equation of
state, both for symmetric-nuclear and pure-neutron matter. Whereas 
$\kappa$ and $\lambda$ are instrumental in the softening of the 
equation of state of symmetric nuclear matter near saturation density, 
$\zeta$ softens the equation of state but at higher densities. 
Finally, the mixed vector coupling 
$\Lambda_{\rm v}$~\cite{Horowitz:2000xj} has been
introduced to soften the density dependence of the symmetry energy 
--- a quantity that is predicted to be stiff in most relativistic 
mean-field models. This effective Lagrangian has been used to compute 
a variety of ground-state observables at the mean-field level and 
will be used here to study the incompressibility of {\sl neutron-rich 
matter}. Further details on the calibration and implementation of the 
relativistic mean-field models may be found in 
Refs.~\cite{Serot:1984ey,Serot:1997xg,Horowitz:2000xj,Todd:2003xs} and
references therein.

Asymmetric nuclear matter is an idealized system consisting of an
infinite number of neutrons and protons interacting exclusively
through the nuclear force. At zero temperature and in the
thermodynamic limit (where both the baryon number $A=N+Z$ and the
volume of the system $V$ tend to infinity) the binding energy per
nucleon depends solely on two intensive variables, the baryon density
$\rho\!\equiv\!A/V$ and the {\sl neutron-proton asymmetry}
$\alpha\!\equiv\!(N-Z)/A$ (the latter may be expressed as
$\alpha\!=\!(\rho_n-\rho_p)/\rho$ in terms of the nucleon densities).
By studying such an idealized system, one hopes to elucidate how the
volume and symmetry terms of the semi-empirical mass
formula~\cite{Bohr:1998} evolve with density.

It has become customary to write the energy per particle of infinite
nuclear matter as follows: 
\begin{equation}
  E/A(\rho,\alpha) - M \equiv {\cal E}(\rho,\alpha)
                          = {\cal E}_{\rm SNM}(\rho)
                          + \alpha^{2}{\cal S}_{2}(\rho)
                          + \alpha^{4}{\cal S}_{4}(\rho)
                          + \ldots\;,
 \label{EOS}
\end {equation}
where we have indicated that $E/A$ is measured relative to the
nucleon rest mass $M$. As the neutron-proton asymmetry is constrained
to the interval $0\!\le\alpha\le\!1$, the total energy per particle
${\cal E}(\rho,\alpha)$ is often expanded in a power series in
$\alpha^{2}$.  Note that odd powers of $\alpha$ do not contribute to
the expansion owing to the symmetry of the strong force between
like-nucleon pairs.  The leading term in this expansion, {\sl i.e.},
${\cal E}_{\rm SNM}(\rho)\!\equiv\!{\cal E}(\rho,\alpha\!=\!0)$,
represents the contribution from symmetric ($N\!=\!Z\!=\!A/2$)
matter. The leading ${\cal O}(\alpha^{2})$-correction to the symmetric
limit, {\sl i.e.}, ${\cal S}_{2}(\rho)\!\equiv\!{\cal S}(\rho)$, is
known as the {\sl symmetry energy}.  The contribution $\alpha^2 {\cal
S}(\rho)$ thus measures the energy involved in converting part of the
protons in symmetric nuclear matter to excess neutrons, at total
baryon density $\rho$.  The above power-series expansion is
particularly useful as the symmetry energy dominates the corrections
to the symmetric limit for all values of $\alpha$. Indeed, to an
excellent approximation the energy per particle of {\sl pure neutron
matter} ($\alpha\!\equiv\!1$) may be written as follows:
%%%
 \begin{equation}
  {\cal E}_{\rm PNM}(\rho)\equiv{\cal E}(\rho,\alpha\!=\!1)
   \approx
  {\cal E}_{\rm SNM}(\rho)+{\cal S}(\rho) \;.
  \label{ApproxPNM} 
 \end{equation}
%%%

The main feature that we aim to understand in the present manuscript
is the evolution with neutron-proton asymmetry of the bulk properties
of infinite nuclear matter --- such as the saturation density, the
binding energy at saturation, and the incompressibility
coefficient. Particularly important is to characterize the sensitivity
of the results to the density dependence of the symmetry energy. To do
so and to establish a baseline, we start by describing the behavior of
symmetric nuclear matter near saturation density.

\subsection{Symmetric Nuclear Matter}
\label{SymmetricMatter} 

One of the hallmarks of the nuclear dynamics is the saturation of
symmetric ($\alpha\!\equiv\!0$) infinite nuclear matter. The
saturation point is characterized by an equilibrium density of about
$\rho_{{}_{0}}\!\simeq\!0.15~{\rm fm}^{-3}$ and an energy per particle
of $\varepsilon_{{}_{0}}\!\simeq\!-16~{\rm MeV}$. Given that the
pressure $P = \rho^{2} \, \partial{\cal E}_{\rm SNM} / \partial\rho$
of symmetric nuclear matter vanishes at saturation, then
the small density fluctuations around the saturation point are fully
characterized by the incompressibility coefficient $K_{0}$. Yet for
reasons that will become clear later, the behavior of symmetric
nuclear matter is expanded in a Taylor series up to third order in the
small density fluctuations. That is,
%%%
\begin{equation}
  {\cal E}_{\rm SNM}(\rho) = \varepsilon_{{}_{0}}+
  \frac{1}{2}K_{0}x^{2}+\frac{1}{6}Q_{0}x^{3} +\ldots\;,
 \label{SNM}
\end{equation} 
%%%
where $x$ is a conveniently defined dimensionless parameter that
characterizes the deviations of the density from its saturation
value, {\sl i.e.,}
%%%
\begin{equation}
  x \equiv \frac{(\rho-\rho_{{}_{0}})}{3\rho_{{}_{0}}} \;.
 \label{XParam}
\end {equation} 
%%%
The various bulk coefficients that characterize the behavior of the
symmetric system near saturation density are given as follows:
%%%
\begin{subequations}
 \begin{align}
  & \varepsilon_{{}_{0}} = {\cal E}_{\rm SNM}(x\!=\!0)
         =  {\cal E}_{\rm SNM}(\rho\!=\!\rho_{{}_{0}})\;,
         \label{ESNM0} \\
  & K_{0} = {\cal E}_{\rm SNM}^{\prime\prime}(x\!=\!0) 
         =9\rho_{{}_{0}}^{2}\left(\frac{\partial^{2}{\cal E}_{\rm SNM}}
         {\partial\rho^{2}}\right)_{\rho=\rho_{{}_{0}}}\;,
         \label{ESNM2} \\
  & Q_{0} = {\cal E}_{\rm SNM}^{\prime\prime\prime}(x\!=\!0) 
         =27\rho_{{}_{0}}^{3}\left(\frac{\partial^{3}{\cal E}_{\rm SNM}}
         {\partial\rho^{3}}\right)_{\rho=\rho_{{}_{0}}}\;.
         \label{ESNM3}
 \end{align}
 \label{Bulk0}
\end{subequations}
%%%

\subsection{Symmetry Energy}
\label{SymmetryEnergy} 

In the so-called {\sl ``parabolic''} approximation the deviations
from the symmetric ($\alpha\!\equiv\!0$) limit are controlled by the 
${\cal O}(\alpha^{2})$-symmetry energy [see Eq.~(\ref{EOS})]. As
has been done for the symmetric case, the behavior of the symmetry
energy around nuclear-matter saturation density may be conveniently 
characterized in terms of a few bulk parameters, namely,
%%%
\begin{equation}
  {\cal S}(\rho) = J + Lx + \frac{1}{2}K_{\rm sym}x^{2}
                 +\frac{1}{6}Q_{\rm sym}x^{3} +\ldots\;,
 \label{SymmE}
\end {equation} 
%%%
where $J$, $L$, $K_{\rm sym}$, and $Q_{\rm sym}$ are the values of the
symmetry energy, slope, curvature, and third derivative at saturation
density. However, unlike symmetric nuclear matter the {\sl ``symmetry
pressure''} $L$ does not vanish. Indeed, the symmetry pressure --- a
quantity that strongly influences the neutron-skin thickness of heavy 
nuclei --- is (within the parabolic approximation) directly
proportional to the pressure of {\sl pure neutron matter}.  That is,
%%%
\begin{equation}
  P_{0} = \frac{1}{3}\rho_{{}_{0}}L\;.
 \label{PZero}
\end {equation} 
%%%

A one-parameter fit to the {\em low-density} behavior of the symmetry
energy that is frequently used in the heavy-ion community is of the
following form~\cite{Tsang:2001jh,Tsang:2004,Shetty:2004jn,
Shetty:2005qp,Shetty:2007zg,Li:2005jy,Li:2008gp}:
%%%
\begin{equation}
 {\cal S}(\rho)={\cal S}_{0}\left(\frac{\rho}
                {\rho_{{}_{0}}}\right)^{\gamma}
               =J(1+3x)^{\gamma} \;,
 \label{SymmE1}
\end{equation}
%%%
where in arriving to the last term we have made use of Eqs.~(\ref{XParam}) 
and~(\ref{SymmE}). To the extent that the above parametrization is accurate, 
the following relations should be satisfied:
%%%
\begin{subequations}
 \begin{align}
  & L=
   \left(\frac{\partial{\cal S}}{\partial x}\right)_{x=0} = 
   3J\gamma\;,
   \label{L} \\
  & K_{\rm sym}= 
   \left(\frac{\partial^{2}{\cal S}}{\partial x^{2}}\right)_{x=0} = 
   9J\gamma(\gamma-1) \;,
   \label{Ksym} \\
  & P_{0}= \rho_{{}_{0}} J\gamma \;.
   \label{Pgamma}
 \end{align}
 \label{BulkSym}
\end{subequations}
%%%
With due caution, mainly because the connection of heavy-ion
experiments to the EOS is not at all trivial and often involves
model-dependent extrapolations of the measured data, significant
constraints on the value of the coefficient $\gamma$ have been
extracted in the last few years from different experimental
observables. For instance, in intermediate-energy heavy-ion
collisions, the analysis of isoscaling data
\cite{Shetty:2005qp,Shetty:2007zg} provides a $\gamma$ value around
0.69. Transport model simulations of data related to isospin diffusion
favor the milder constraint $\gamma \sim 0.69$--$1.05$
\cite{Li:2005jy,Li:2008gp}.  Some nuclear collective modes
provide another tool to probe the behavior of ${\cal S}(\rho)$ at
subsaturation densities. The values $P_{0}=2.3\pm0.8$ MeV/fm$^3$ and
$J=32\pm1.8$ MeV extracted in Ref.~\cite{Klimkiewicz:2007zz} from
pygmy dipole resonances suggest a value of $\gamma\sim0.5\pm0.15$,
whereas the constraint $23.3 < {\cal S}(\rho=0.1 \, {\rm fm}^{-3}) <
24.9$ MeV obtained in Ref.~\cite{Trippa:2008gr} from the properties of
the giant dipole resonance in $^{208}$Pb hints at a value of $\gamma
\sim 0.5$--0.65. These findings from experimental isospin-sensitive
signals imply a rather soft nuclear symmetry energy at subsaturation
densities. An analysis \cite{Centelles:2008vu} of a set of neutron
skins of nuclei measured across the mass table by antiprotonic
techniques yields a similar conclusion. Finally, recent studies 
of the low-density behavior of pure neutron matter using universal
properties of dilute Fermi gases seem to support the same 
findings~\cite{Schwenk:2005ka,Piekarewicz:2007dx}.

\subsection{Neutron-Rich Matter}
\label{NeutronRichMatter} 

Insofar as neutron-rich matter saturates, the energy per particle may
continue to be written as in Eq.~(\ref{SNM}). Thus, the aim of this
section is to characterize the evolution of the saturation point---in
particular, the saturation density, the binding energy per nucleon, and 
the incompressibility coefficient---as a function of the neutron-proton 
asymmetry $\alpha\!=\!(N-Z)/A$. To do so, it will prove instructive to 
proceed along two alternative paths: one purely analytical and the other
purely numerical. In the analytic case, the saturation properties of 
neutron-rich matter will be derived from a handful of bulk 
parameters that characterize the behavior of both symmetric nuclear 
matter and the symmetry energy around saturation density, as was done
in Eqs.~(\ref{SNM}) and~(\ref{SymmE}). This purely analytic procedure, 
already well known in the literature, will be contrasted against a 
numerical procedure that is free of any assumptions or approximations
beyond that of the mean-field approximation. We will verify that these
two alternative approaches agree at the few percent level, thereby
lending support to the analytic approach in elucidating the evolution 
of the incompressibility coefficient with neutron excess.

According to Eqs.~(\ref{SNM}) and~(\ref{SymmE}), the energy per particle 
of asymmetric nuclear matter with a neutron-proton asymmetry $\alpha$ 
may be written in the form
%%%
\begin{equation}
 {\cal E}(\rho,\alpha) \approx 
 (\varepsilon_{{}_{0}}+J\alpha^{2})+L\alpha^{2}x + 
 \frac{1}{2}(K_{0}+\alpha^{2}K_{\rm sym})x^{2}+
 \frac{1}{6}(Q_{0}+\alpha^{2}Q_{\rm sym})x^{3}\;.
 \label{EvsX}
\end{equation}
%%%
Clearly, the presence of the linear term shifts the saturation
point from $x_{0}\!\equiv\!0$ to $\bar{x}_{0}$, where the latter
is defined as the physical solution to the following equation:
%%%
\begin{equation}
 \left(\frac{\partial{\cal E}}{\partial x}\right)=
 \alpha^{2}L + (K_{0}+\alpha^{2}K_{\rm sym})x
       + \frac{1}{2}(Q_{0}+\alpha^{2}Q_{\rm sym})x^{2}=0\;.
 \label{NewX}
\end{equation}
%%%
Although the roots of this equation may be found by solving a simple
quadratic equation, the ${\cal O}(\alpha^{2})$ solution may be
solved by inspection. One obtains
%%%
\begin{equation}
  \bar{x}_{{}_{0}}=-\frac{L}{K_{0}}\alpha^{2}+{\cal O}(\alpha^{4})
  \quad{\rm or}\quad
  {\overline{\rho}_{{}_{0}}}/\rho_{{}_{0}} = 1+3\bar{x}_{{}_{0}}=
  1-3\frac{L}{K_{0}}\alpha^{2}+{\cal O}(\alpha^{4})\;.
 \label{X0Bar}
\end{equation}
%%%
The values for the energy per particle and the incompressibility
coefficient may now be found by expanding Eq.~(\ref{EvsX}) around 
this new ($\alpha\!\ne\!0$) value of the saturation density:
%%%
\begin{equation}
 {\cal E}(\rho,\alpha) = {\cal E}(\bar{x}_{{}_{0}},\alpha)
 +\frac{1}{2}(x-\bar{x}_{{}_{0}})^{2}{\cal E}''(\bar{x}_{{}_{0}},\alpha)
 +\ldots\;.
 \label{NewEOS}
\end{equation}
Alternatively, by introducing the dimensionless parameter 
%%%
\begin{equation}
  \bar{x} \equiv \frac{(\rho-\bar{\rho}_{{}_{0}})}{3\bar{\rho}_{{}_{0}}}
 \label{XBarParam}
\end {equation} 
%%%
to characterize the deviations of the density from the {\sl new
saturation point}, the expansion of ${\cal E}(\rho,\alpha)$ given in
Eq.~(\ref{NewEOS}) may be cast in a form analogous to Eq.~(\ref{SNM})
for ${\cal E}_{\rm SNM}(\rho)$. That is,
%%%
\begin{equation}
 {\cal E}(\rho,\alpha) = {\cal E}(\bar{x}_{{}_{0}},\alpha)
 +\frac{1}{2}\Big[(1+3\bar{x}_{{}_{0}})^2
 {\cal E}''(\bar{x}_{{}_{0}},\alpha)\Big]\bar{x}^{2}
 +\ldots \equiv
 \varepsilon_{{}_{0}}(\alpha)+\frac{1}{2}K_{0}(\alpha)\bar{x}^{2}
 +\ldots\;,
 \label{SNMlike}
\end{equation}
%%%
where the energy per particle and the incompressibility coefficient
at the new saturation density are given by
%%%
\begin{subequations}
 \begin{align}
 &  \varepsilon_{{}_{0}}(\alpha)\equiv{\cal E}(\bar{x}_{{}_{0}},\alpha)
   =\varepsilon_{{}_{0}}+J\alpha^{2}+{\cal O}(\alpha^{4}) \;, \\
 & K_{0}(\alpha)\equiv (1+3\bar{x}_{{}_{0}})^2
   {\cal E}''(\bar{x}_{{}_{0}},\alpha)=
   K_{0}+\Big(K_{\rm sym}-6L-\frac{Q_{0}}{K_{0}}L\Big)\alpha^{2}
   +{\cal O}(\alpha^{4}) \;.
 \end{align}
 \label{NewEandK}
\end{subequations}
%%%

The analytic results correct to second order in $\alpha$ are summarized 
in the following set of equations where the quantities $\rho_{\tau}$,
$\varepsilon_{\tau}$, and $K_{\tau}$ represent the {\em deviations} of the
saturation density, energy per particle, and incompressibility coefficient
of \textsl{infinite matter} away from the symmetric $N\!=\!Z$ limit:
%%%
\begin{subequations}
 \begin{align}
  & \rho_{{}_{0}}(\alpha)=\rho_{{}_{0}} + \rho_{\tau}\alpha^{2} 
    + {\cal O}(\alpha^{4})
    = \rho_{{}_{0}}-3\rho_{{}_{0}}\frac{L}{K_{0}}\alpha^{2} 
    + {\cal O}(\alpha^{4}) \;,
   \label{RhoTau} \\
  & \varepsilon_{{}_{0}}(\alpha)=\varepsilon_{{}_{0}}
               +\varepsilon_{\tau}\alpha^{2} + {\cal O}(\alpha^{4})
    =\varepsilon_{{}_{0}}+J\alpha^{2} + {\cal O}(\alpha^{4}) \;,
   \label{ETau} \\
  & K_{0}(\alpha)=K_{0}+K_{\tau}\alpha^{2}+{\cal O}(\alpha^{4})
  = K_{0}+\Big(K_{\rm sym}-6L-\frac{Q_{0}}{K_{0}}L\Big)\alpha^{2}
   +{\cal O}(\alpha^{4}) \;.
 \label{KTau}
 \end{align}  
 \label{Taus}
\end{subequations}
%%%
In view of the profuse choices of terminology existing in the literature,
our notation and conventions are discussed further in the Appendix.

\section{Results}
\label{Results} 

Having developed the necessary formalism in the preceding sections, we
devote this section to present the results of our calculations.  As we
have done elsewhere~\cite{Piekarewicz:2007us}, our results were
generated using two accurately calibrated models:
NL3~\cite{Lalazissis:1996rd,Lalazissis:1999} and
FSUGold~\cite{Todd-Rutel:2005fa}. In addition, we performed
calculations with a Hybrid model to be introduced later. Effective
meson masses ({\sl i.e.,} interaction ranges) and coupling constants
for the present models are displayed in Table~\ref{Table1} as defined
by the Lagrangian density of Eq.~(\ref{Lagrangian}).

%%%%%%%%%%%%%%%%%%%%%%%%%%%%%%%%%%%%%%%%%%%%%%%%%%%%%%%%%%%%%%%%%
\begin{table}
\begin{tabular}{|l||c|c|c|c|c|c|c|c|c|c|}
 \hline
 Model & $m_{\rm s}$  & $m_{\rm v}$  & $m_{\rho}$  
       & $g_{\rm s}^2$ & $g_{\rm v}^2$ & $g_{\rho}^2$
       & $\kappa$ & $\lambda$ & $\zeta$ & $\Lambda_{\rm v}$\\
 \hline
 \hline
 FSU    & 491.500 & 782.500 & 763.000 & 112.1996 & 204.5469 & 138.4701 
        & 1.4203  & $+$0.02376 & 0.06 & 0.03 \\
 NL3    & 508.194 & 782.501 & 763.000 & 104.3871 & 165.5854 &  79.6000 
        & 3.8599  & $-$0.01591 & 0.00 & 0.00 \\
 Hybrid & 508.194 & 782.501 & 763.000 & 106.2575 & 165.5848 &  79.6483
        & 4.5472  & $-$0.01952 & 0.00 & 0.00 \\
 \hline
\end{tabular}
\caption{Model parameters used in the calculations. The parameter
$\kappa$ and the meson masses $m_{\rm s}$, $m_{\rm v}$, and $m_{\rho}$
are all given in MeV\@. The value of the nucleon mass is taken as
$M\!=\!939$~MeV.} 
\label{Table1}
\end{table}
%%%%%%%%%%%%%%%%%%%%%%%%%%%%%%%%%%%%%%%%%%%%%%%%%%%%%%%%%%%%%%%%%

With the above sets of parameters, one may compute the nuclear-matter
equation of state, namely, the energy per particle as a function of
density and neutron excess. In particular, one can extract values for
the various bulk parameters defined in Eqs.~(\ref{SNM})
and~(\ref{SymmE}) that characterize the behavior of neutron-rich
matter around saturation density; these parameters are listed in
Table~\ref{Table2}. Note that in order to make contact with the
parametrization given in Eq.~(\ref{SymmE1}), the value of the exponent
$\gamma$ listed in Table~\ref{Table2} was extracted from a fit to the
symmetry energy in the low-density range of
$\rho\!=\!(0.3\!-\!1.0)\rho_{{}_{0}}$. The found results are in
consonance with the prediction $\gamma=L/3J$ that follows from
Eq.~(\ref{SymmE1}) (namely, $\gamma=0.62$ for FSUGold and
$\gamma=1.06$ in the case of NL3 and the Hybrid model). Whereas the
FSUGold and NL3 models agree on the energy and density at
saturation---quantities that are tightly constrained by existent
ground-state observables---significant discrepancies emerge in all
remaining parameters. The main difference between the two models may
be succinctly summarized by stating that whereas FSUGold predicts a
soft behavior for both symmetric nuclear matter (through $K_{0}$) and
the symmetry energy (through $L$), NL3 predicts a stiff behavior for
both. Note that {\sl ``stiff''} or {\sl ``soft''} refers to whether
the energy increases rapidly or slowly with density.

%%%%%%%%%%%%%%%%%%%%%%%%%%%%%%%%%%%%%%%%%%%%%%%%%%%%%%%%%%%%%%%%%
\begin{table}
\begin{tabular}{|l||c|c|c|c|c|c|c|c|}
 \hline
 Model & $\rho_{0}$ & $\varepsilon_{0}$
       & $K_{0}$ & $Q_{0}$ & $J$ & $L$ 
       & $K_{\rm sym}$  & $\gamma$\\
 \hline
 \hline
 FSU     &  0.148   & $-$16.30 & 230.0 & $-$523.4 & 32.59 &  60.5
         & $-$51.3  & 0.64 \\
 NL3     &  0.148   & $-$16.24 & 271.5 & $+$204.2 & 37.29 & 118.2
         & $+$100.9 & 0.98 \\
 Hybrid  &  0.148   & $-$16.24 & 230.0 & $-$71.5  & 37.30 & 118.6
         & $+$110.9 & 0.98 \\
\hline
\end{tabular}
\caption{Bulk parameters (as described in the text) characterizing 
         the energy of symmetric nuclear matter [Eq.~(\ref{SNM})] 
         and the symmetry energy [Eq.~(\ref{SymmE})] at saturation 
         density. 
         All quantities are in MeV, with the exception of $\rho_{0}$ 
         given in fm$^{-3}$ and the dimensionless parameter $\gamma$ 
         defined in Eq.~(\ref{SymmE1}).} 
\label{Table2}
\end{table}
%%%%%%%%%%%%%%%%%%%%%%%%%%%%%%%%%%%%%%%%%%%%%%%%%%%%%%%%%%%%%%%%%

That symmetric nuclear matter and the symmetry energy are either both
soft (as in the FSUGold model) or both stiff (as in the NL3 model) may
lie at the core of the problem in reproducing the experimentally
measured GMR energies in the
Sn-isotopes~\cite{Garg:2006vc,Li:2007bp,Piekarewicz:2007us}.
According to Eq.~(\ref{KTau}), a large value of $L$ (as in NL3)
generates a large softening of the incompressibility coefficient
relative to its value in symmetric nuclear matter. However, a large
incompressibility coefficient in symmetric nuclear matter $K_{0}$ (as
in NL3) hinders the softening generated by $K_{\tau}$. Conversely,
FSUGold predicts a relatively small value for $K_{0}$. However, its
soft symmetry energy generates a small (absolute) value for $K_{\tau}$
that precludes the significant reduction in the incompressibility
coefficient required by the experimental GMR energies. In an effort to
circumvent this problem---and this problem only---we have generated a
Hybrid model having the same incompressibility coefficient as FSUGold
while preserving the stiff symmetry energy of NL3 (see
Table~\ref{Table2}).  As seen in Table~\ref{Table1}, this was
accomplished through a slight adjustment of the scalar self-coupling
parameters $\kappa$ and $\lambda$. 
%%%%%%% ADDED IN RESPONSE TO THE REFEREE'S COMMENTS %%%%%%%
Note that it is not our intent to accurately calibrate the Hybrid model 
introduced here. Rather, the Hybrid model --- although reasonably 
accurate --- should merely be regarded as a {\sl ``test''} model
that illustrates how surprisingly soft are the experimental GMR energies of
the Tin isotopes relative to the theoretical predictions. Indeed, as we
will display later in Fig.~\ref{Fig8}, not even such an artificially-tuned 
model can fully account for the rapid softening of the GMR energies
with neutron excess. Let us mention that
%%%%%%%%%%%%%%%%%%%%%%%%%%%%%%%%%%%%%%%%%%%%%%%%%%%
nowadays it is generally acknowledged that {\sl experimental
data} on compressional-mode giant resonances support a value of
$K_{0}\!\approx\!240$~MeV~\cite{Youngblood:1999} for the
incompressibility coefficient of symmetric nuclear matter. Recently
measured data on the breathing mode of Sn isotopes seem to favor a
constraint $K_{\tau}\!=\!-550\!\pm\!100$~MeV for the asymmetry term in
the nuclear incompressibility~\cite{Garg:2006vc,Li:2007bp}. A similar
value of $K_{\tau}\!\sim-\!500$~MeV has been reported from independent
experimental evidence available from isospin diffusion observables in
nuclear reactions \cite{Li:2005jy,Li:2008gp} and from neutron skins of
nuclei~\cite{Centelles:2008vu}. Tables~\ref{Table2} and~\ref{Table3}
confirm that the Hybrid model is consistent with both of the indicated
$K_{0}$ and $K_{\tau}$ values.

%%%%%%%%%%%%%%%%%%%%%%%%%%%%%%%%%%%%%%%%%%%%%%%%%%%%%%%%%%%%%%%%%%%%%%
\begin{figure}[ht]
\vspace{0.50in}
\includegraphics[height=4in,angle=0]{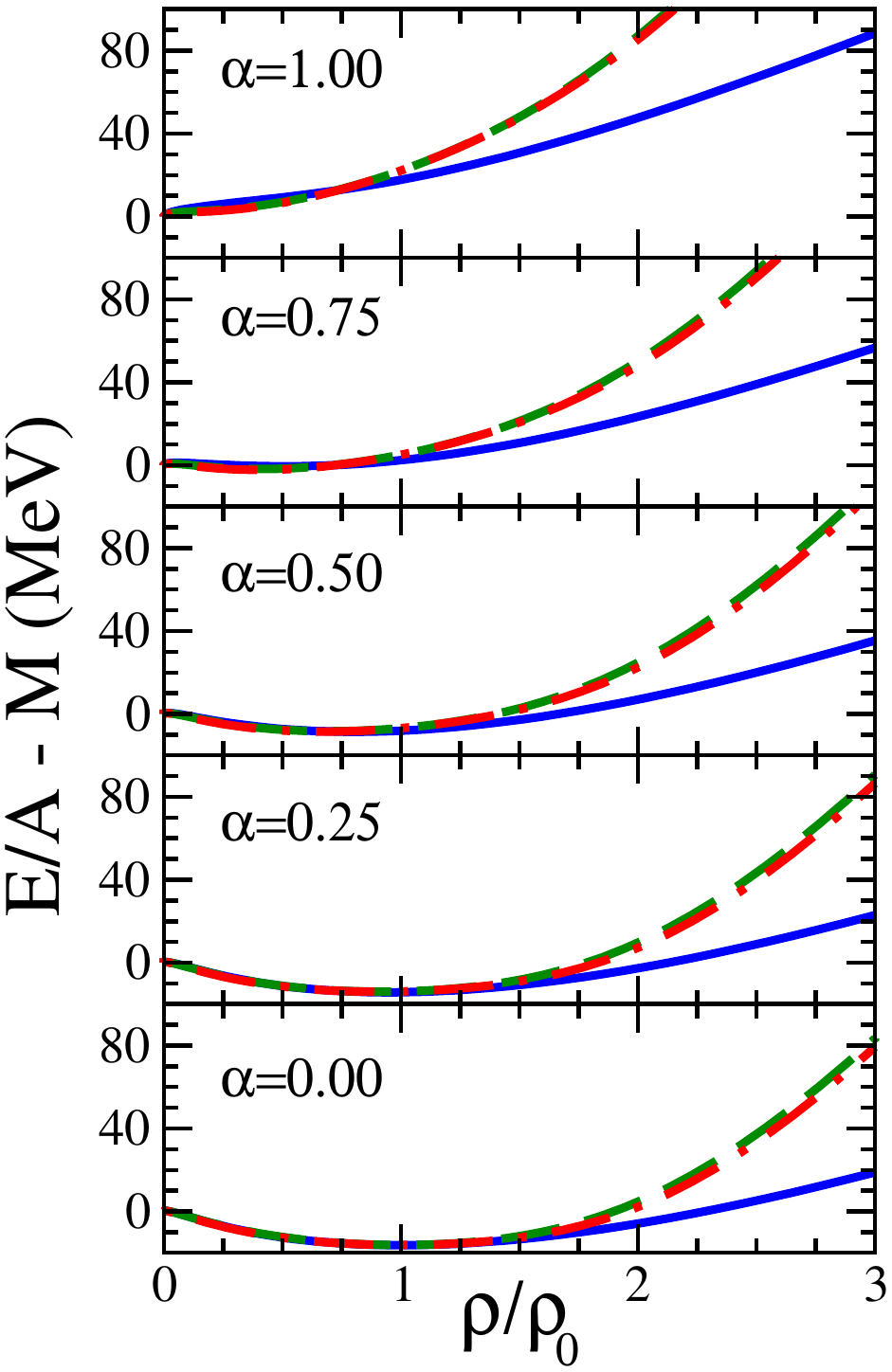}
\caption{(Color online) 
Density dependence of the energy per particle in infinite nuclear
matter at the shown neutron-proton asymmetries according to the
FSUGold (blue solid line), NL3 (green dashed line), and Hybrid (red
dot-dashed line) models.}
\label{Fig1}
\end{figure}
%%%%%%%%%%%%%%%%%%%%%%%%%%%%%%%%%%%%%%%%%%%%%%%%%%%%%%%%%%%%%%%%%%%%%%

\begin{figure}[ht]
\vspace{0.50in}
\includegraphics[width=5in,angle=0]{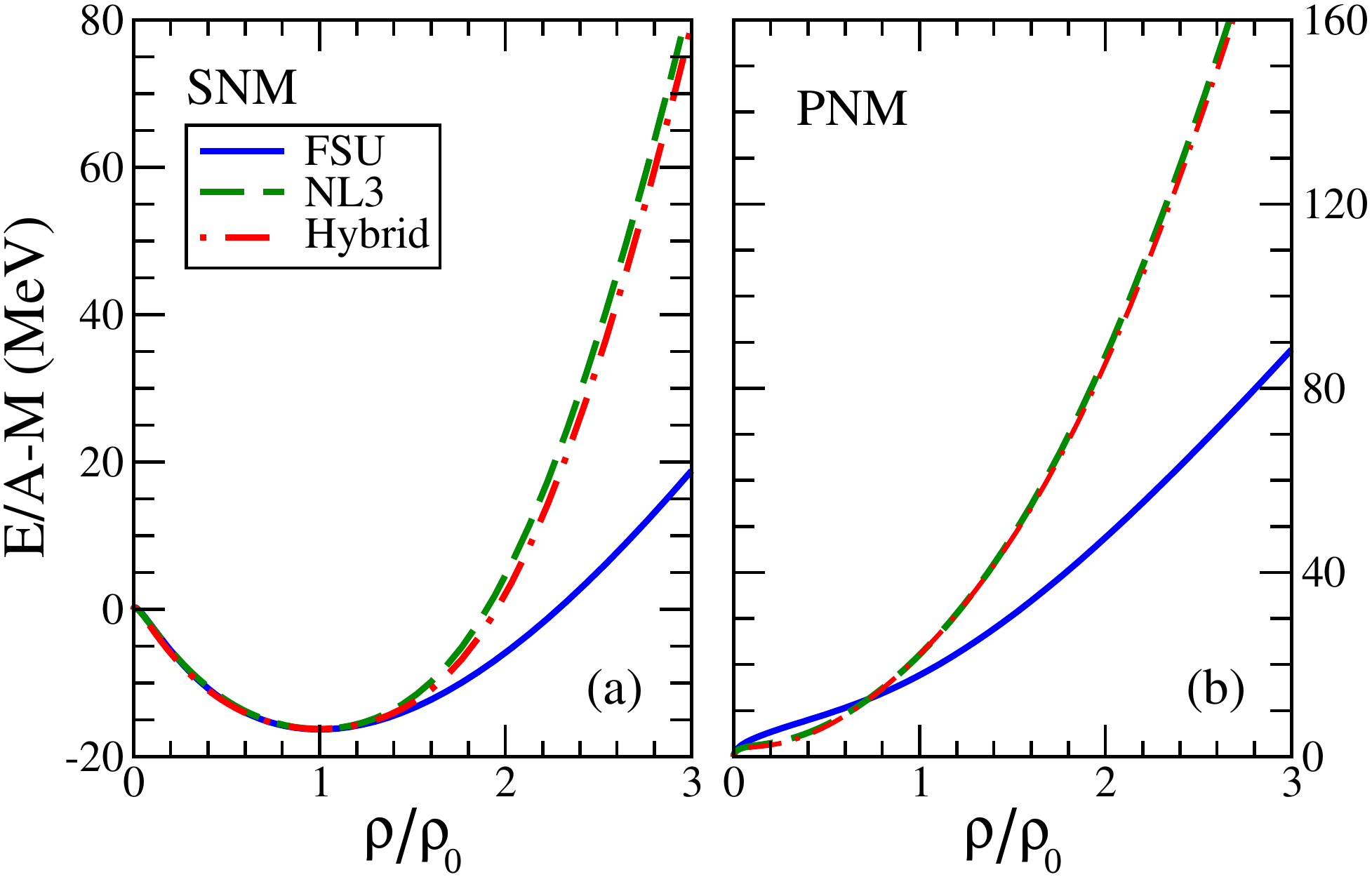}
\caption{(Color online)
Density dependence of the energy per particle in symmetric nuclear matter
(SNM) and in pure neutron matter (PNM) for the investigated models.}
\label{Fig2}
\end{figure}

\begin{figure}[ht]
\vspace{0.50in}
\includegraphics[height=4in,angle=0]{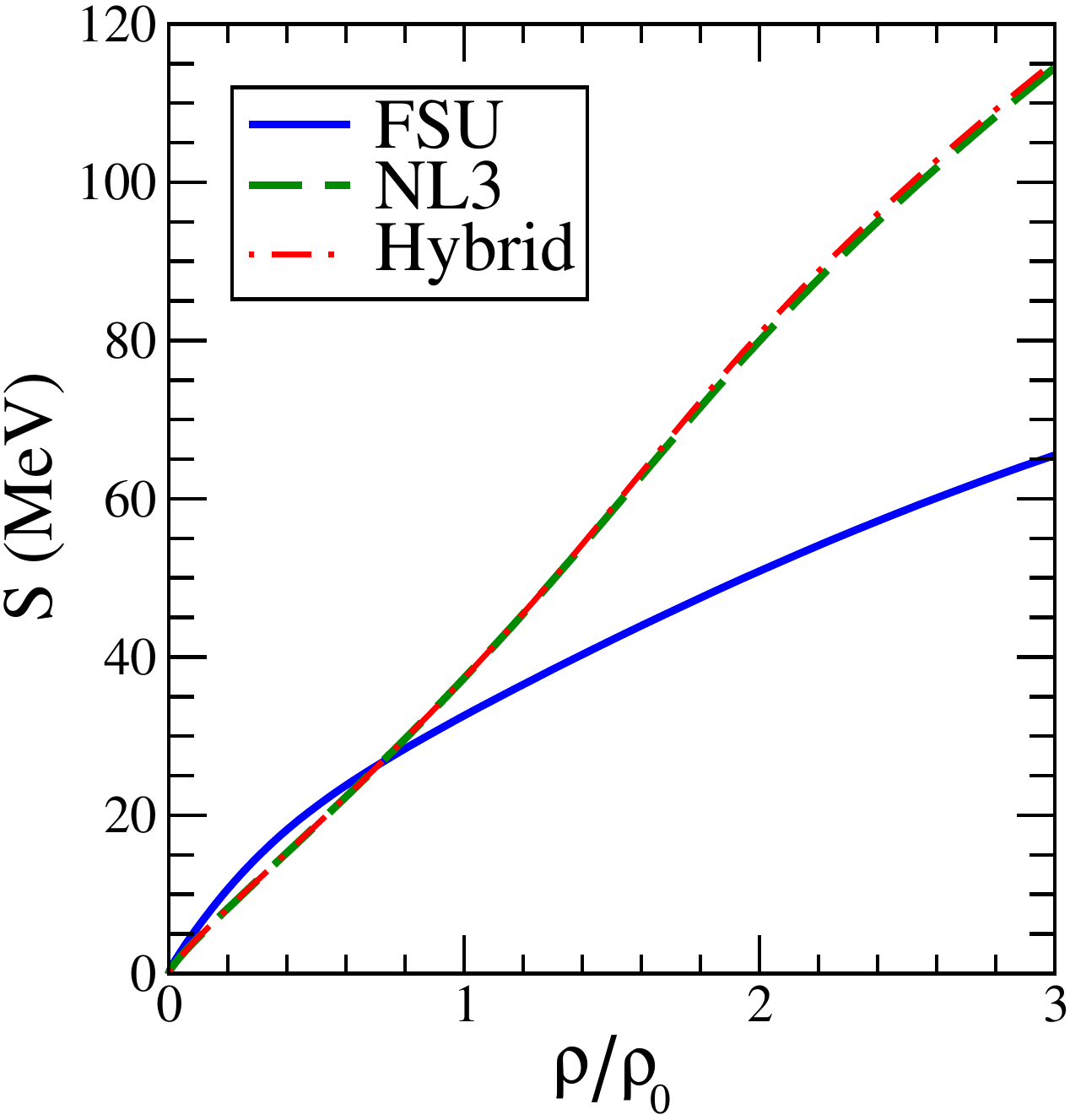}
\caption{(Color online)
Density dependence of the symmetry energy coefficient for the
investigated models.}
\label{Fig3}
\end{figure}

The evolution of the equation of state with neutron-proton asymmetry
$\alpha$ is displayed in Fig.~\ref{Fig1} for the three models
considered in the text: FSUGold (solid blue lines), NL3 (dashed green
lines), and Hybrid (dot-dashed red lines). The $\alpha\!=\!0$ curve
corresponds to symmetric nuclear matter whereas the $\alpha\!=\!1$
curve corresponds to pure neutron matter. In all models nuclear matter
ceases to saturate at a value of the neutron-proton asymmetry slightly
larger than $\alpha\!=\!0.75$. Figure~\ref{Fig2} provides an expanded
version of the results for symmetric nuclear matter ($\alpha\!=\!0$)
and for pure neutron matter ($\alpha\!=\!1$). To an excellent
approximation---especially at the subsaturation densities of
relevance to this work---the difference between the equation of state
of pure neutron matter and that of symmetric matter equals the
symmetry energy of Fig.~\ref{Fig3}. 
On the other hand, a truncated expansion ${\cal S}(\rho) = J + Lx +
\frac{1}{2}K_{\rm sym}x^{2}$ [cf.\ Eq.~(\ref{SymmE})] appears to
provide a fair enough representation of the actual value of ${\cal
S}(\rho)$ in a range of densities roughly between half and twice the
saturation density of symmetric nuclear matter. Indeed, we find that
the discrepancies of this approximation compared with the exact ${\cal
S}(\rho)$ are less than 5\% in a density range $0.45\rho_{{}_{0}}
\lesssim \rho \lesssim 2.7\rho_{{}_{0}}$ for FSUGold and
$0.33\rho_{{}_{0}} \lesssim \rho \lesssim 2.15\rho_{{}_{0}}$ for NL3
and the Hybrid model.

One observes that in all cases the Hybrid model seems to follow
closely the predictions of the NL3 model in
Figs.~\ref{Fig1}--\ref{Fig3}. For the symmetry energy this has been
done by construction. For symmetric nuclear matter, however, the
Hybrid model is indeed softer than NL3---and as soft as FSUGold---at
saturation density. That the Hybrid model tracks NL3 at {\sl high
density} is a reflection of the vector self-coupling parameter $\zeta$
having been set to zero in both models. This confirms that the value
of the incompressibility coefficient of symmetric nuclear matter at
saturation density has practically no impact on the equation of state
of high-density matter and, by extension, on most neutron-star
properties~\cite{Mueller:1996pm}. 

%%%%%%%%%%%%%%%%%%%%%%%%%%%%%%%%%%%%%%%%%%%%%%%%%%%%%%%%%%%%%%%%%%%%%%
%%%%%%%%%%%%%%%%%%%%%%%%%%%%%%%%%%%%%%%%%%%%%%%%%%%%%%%%%%%%%%%%%%%%%%

From the evolution of the equation of state with $\alpha$ we conclude
quite generally that as the neutron-proton asymmetry increases, the
saturation density is reduced, the binding energy weakens, and the
nuclear incompressibility softens. Based on the structure of
Eqs.~(\ref{Taus}), we regard these trends as model independent.
Indeed, that the saturation density is reduced ({\it i.e.,}
$\rho_{\tau}\!<\!0$) is a simple reflection of (a) symmetric nuclear
matter has a stable equilibrium point ($K_{0}\!>\!0$) and (b) the
pressure of pure neutron matter at saturation density is positive
($L\!>\!0$). Further, that the binding energy weakens ({\it i.e.,}
$\varepsilon_{\tau}\!>\!0$) follows from the fact that pure neutron
matter is not self bound, namely, $J\!\ge\!|\epsilon_{0}|$. Finally,
that the nuclear incompressibility softens requires $K_{\tau}\!<\!0$.
This is the hardest condition to satisfy as it depends on higher-order
derivatives of the equation of state, namely, on $Q_{0}$ and $K_{\rm
sym}$ [see Eq.~(\ref{KTau})]. However, barring anomalously large
values for these two quantities, the condition $K_{\tau}\!<\!0$ hinges
also on the pressure of pure neutron matter at saturation density
being positive. This is due to the large coefficient multiplying $L$
in Eq.~(\ref{KTau}), which provides the dominant contribution to
$K_{\tau}$ as compared to $K_{\rm sym}$ and $Q_{0}/K_{0}$.  We
conclude that whereas the signs of $\rho_{\tau}$,
$\varepsilon_{\tau}$, and $K_{\tau}$ are fairly model independent,
their model-dependent magnitude is determined by two fundamental
parameters of the equation of state: the incompressibility coefficient
of symmetric nuclear matter $K_{0}\!>\!0$ and the symmetry
pressure~$L$.

%%%%%%%%%%%%%%%%%%%%%%%%%%%%%%%%%%%%%%%%%%%%%%%%%%%%%%%%%%%%%%%%%%%%%%
%%%%%%%%%%%%%%%%%%%%%%%%%%%%%%%%%%%%%%%%%%%%%%%%%%%%%%%%%%%%%%%%%%%%%%

To examine the evolution of the saturation point with neutron-proton
asymmetry we have tabulated in Table~\ref{Table3} the values for
$\rho_{\tau}$, $\varepsilon_{\tau}$, and $K_{\tau}$.  These quantities
(which have been enclosed in parenthesis) were computed directly from
the analytic expressions given in Eqs.~(\ref{Taus}). The bulk
parameters that they depend on were previously extracted from a fit to 
the equation of state of symmetric nuclear matter and to the symmetry
energy and are listed in Table~\ref{Table2}. As alluded earlier, the
fact that neither $Q_{0}$ nor $K_{\rm sym}$ are anomalously large in
the present models, results in an asymmetry term in the nuclear
incompressibility $K_{\tau}$ that is dominated by the symmetry
pressure $L$. In particular, note that the value of $K_{\tau}$ in the
Hybrid model is consistent with the value extracted in
Refs.~\cite{Garg:2006vc,Li:2007bp} from the measurement of the GMR
energies in the Sn-isotopes.

%
% FSU    & $-$0.117 ($-$0.117) & 32.60 (32.59) & $-$275.45 ($-$275.66)\\
% NL3    & $-$0.188 ($-$0.194) & 37.24 (37.29) & $-$682.65 ($-$698.08)\\
% Hybrid & $-$0.215 ($-$0.229) & 37.20 (37.30) & $-$531.98 ($-$567.12)\\
%

%%%%%%%%%%%%%%%%%%%%%%%%%%%%%%%%%%%%%%%%%%%%%%%%%%%%%%%%%%%%%%%%%
\begin{table}
\begin{tabular}{|l||c|c|c|}
 \hline
 Model & $\rho_{\tau}({\rm fm}^{-3})$ & $\varepsilon_{\tau}$(MeV) 
       & $K_{\tau}$(MeV) \\
 \hline
 \hline
 FSU    & $-$0.117 ($-$0.117) & 32.60 (32.59) & $-$275.45 ($-$276.77)\\
 NL3    & $-$0.188 ($-$0.194) & 37.24 (37.29) & $-$682.65 ($-$697.36)\\
 Hybrid & $-$0.215 ($-$0.229) & 37.20 (37.30) & $-$531.98 ($-$563.86)\\
\hline
\end{tabular}
\caption{Leading ${\cal O}(\alpha^{2})$-correction to the 
         evolution of the saturation density, energy per 
         particle, and incompressibility coefficient of
         asymmetric nuclear matter. Quantities outside
         the parentheses were extracted from a quadratic
         fit to the numerical results in the 
         $0\!\le\!\alpha^{2}\!\le\!0.1$ range, whereas
         the quantities in parenthesis were
         computed from the analytic expressions given in
         Eqs.~(\ref{Taus}).}
\label{Table3}
\end{table}
%%%%%%%%%%%%%%%%%%%%%%%%%%%%%%%%%%%%%%%%%%%%%%%%%%%%%%%%%%%%%%%%%

\begin{figure}[ht]
\vspace{0.50in}
\includegraphics[width=5in,angle=0]{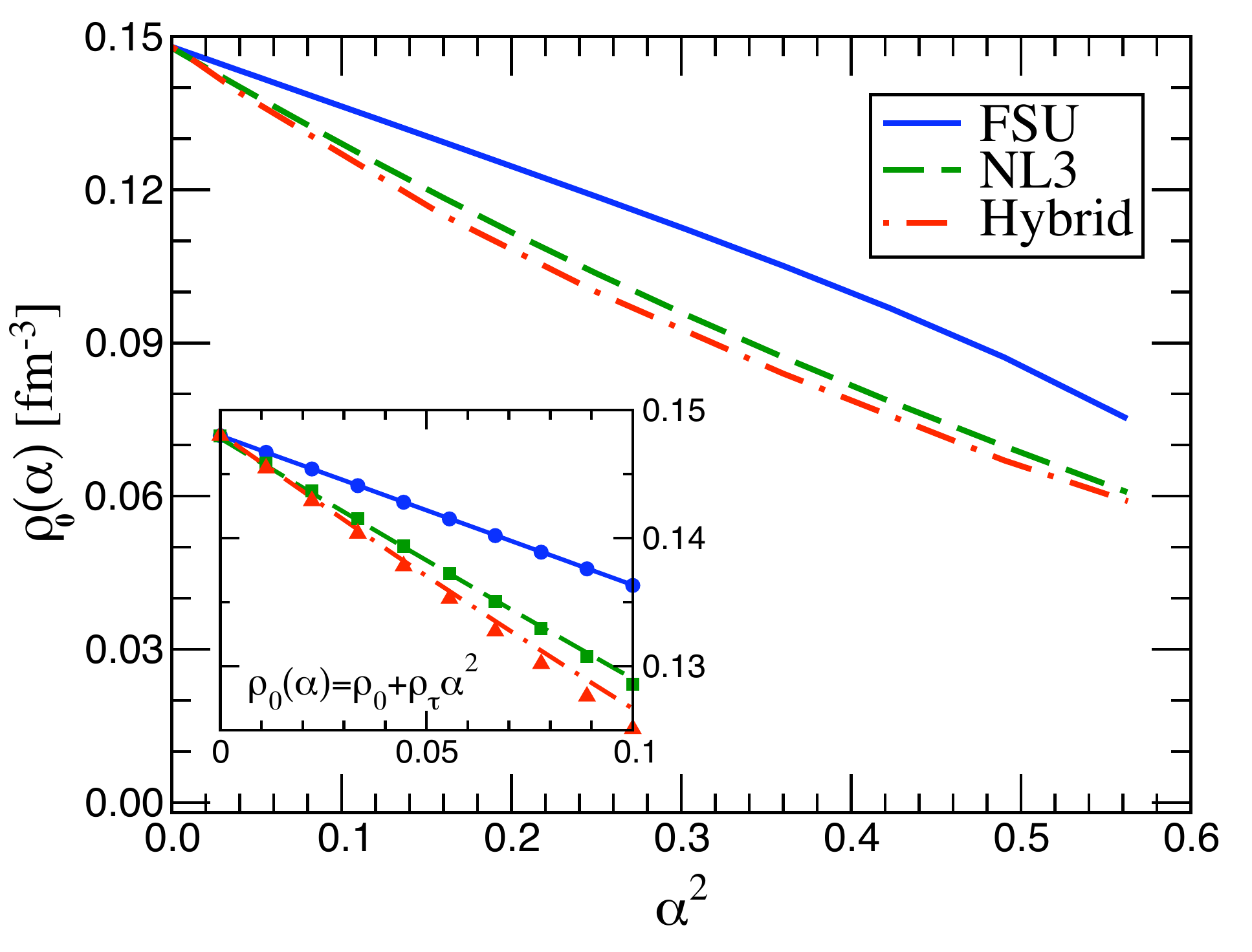}
\caption{(Color online) 
Evolution with increasing neutron-proton asymmetry of the baryon
density that corresponds to the exact saturation point of asymmetric
nuclear matter. The inset displays by symbols the result of a
least-squares fit in the $0\!\le\!\alpha^{2}\!\le\!0.1$ range assuming
a parabolic dependence.}
\label{Fig4}
\end{figure}

\begin{figure}[ht]
\vspace{0.50in}
\includegraphics[width=5in,angle=0]{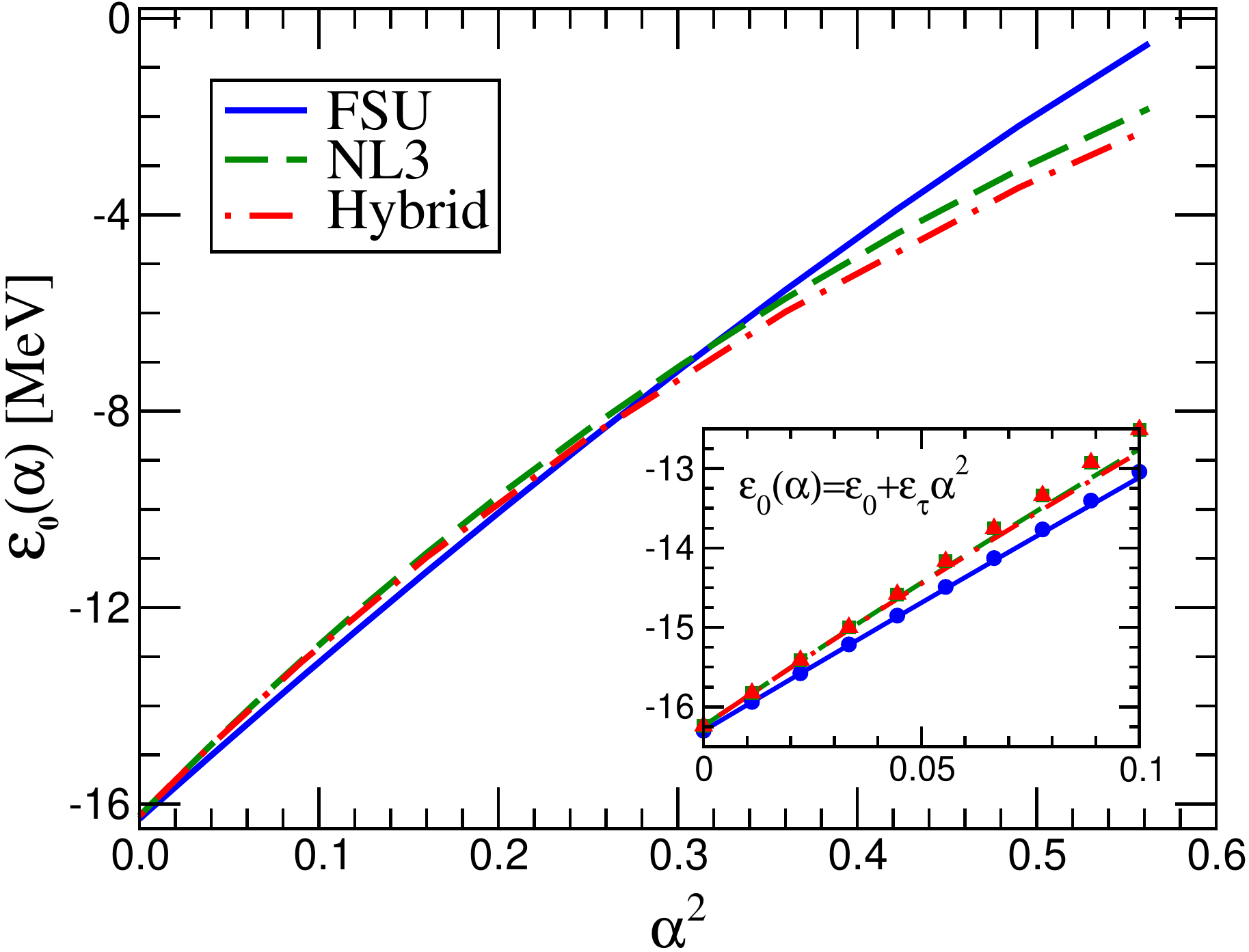}
\caption{(Color online)
Same as described in the caption to Fig.~\ref{Fig4} for the energy per
particle.}
\label{Fig5}
\end{figure}

\begin{figure}[ht]
\vspace{0.50in}
\includegraphics[width=5in,angle=0]{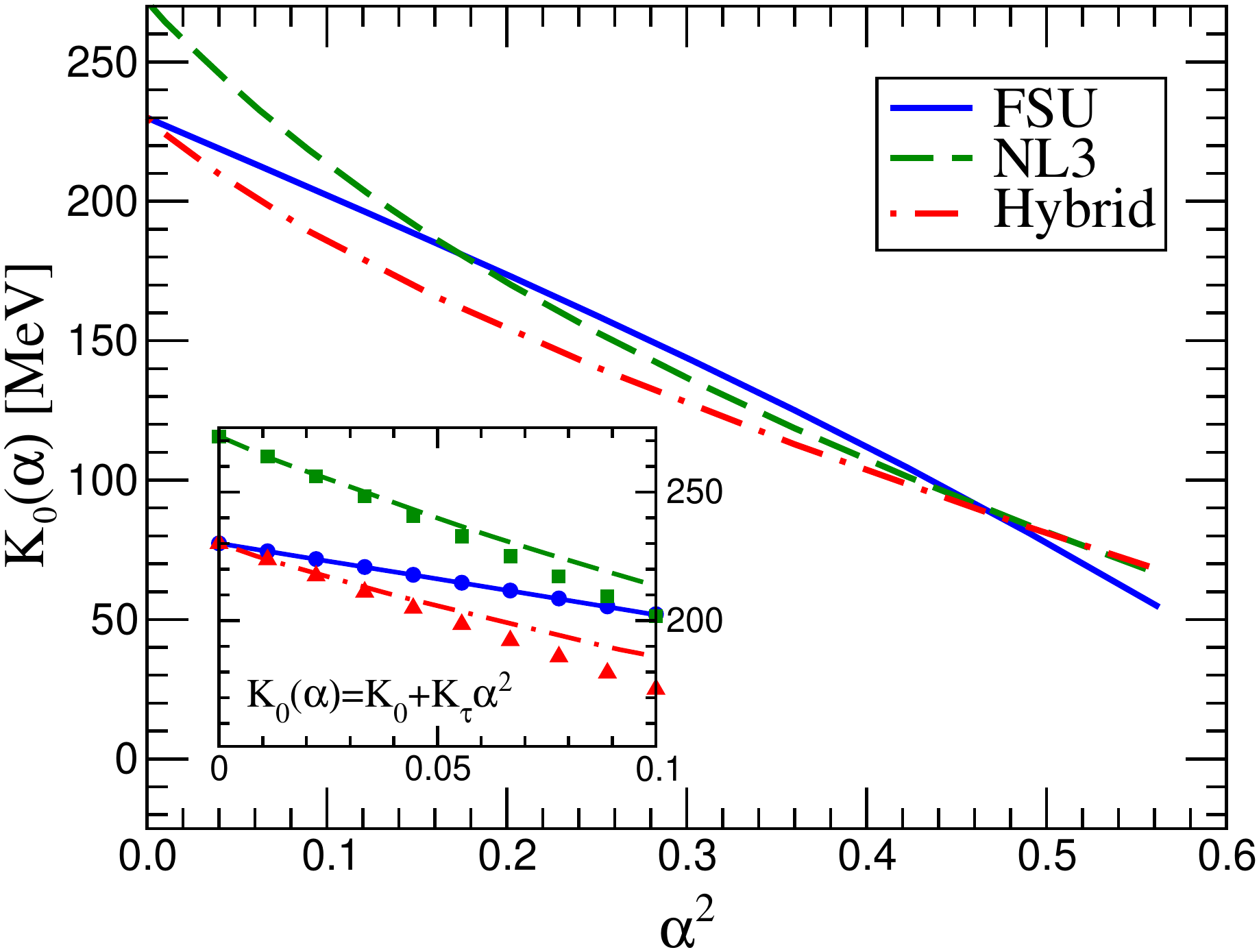}
\caption{(Color online)
Same as described in the caption to Fig.~\ref{Fig4} for the
incompressibility coefficient.}
\label{Fig6}
\end{figure}

But how accurate are the expressions given in Eqs.~(\ref{Taus})? To
test the reliability of these analytic expressions we have carried out
a purely numerical exercise that is exact within the purview of the
mean-field approximation. There is no reliance on the parabolic
approximation, as in Eq.~(\ref{EOS}), or on expansions around the
saturation density of symmetric matter, such as in Eq.~(\ref{EvsX}).
Basically, the equation of state of asymmetric nuclear
matter is computed numerically for a range of values of the
neutron-proton asymmetry in the $0\!\le\!\alpha^{2}\!\le\!0.1$
range. For each value of $\alpha$, the new saturation point---namely
the density, energy-per-nucleon, and incompressibility coefficient at
the minimum---is computed.  Once this procedure is completed, one
extracts the three desired coefficients ($\rho_{\tau}$,
$\varepsilon_{\tau}$, and $K_{\tau}$) from a least-squares fit to the
numerical data. Such a procedure is illustrated in
Figs.~\ref{Fig4},~\ref{Fig5}, and \ref{Fig6} for the saturation
density, energy-per-particle, and incompressibility coefficient,
respectively. In all cases the inset includes a comparison between the
exact numerical results (displayed with lines) and the analytic
approximations (displayed with symbols). Moreover, $\rho_{\tau}$,
$\varepsilon_{\tau}$, and $K_{\tau}$ have also been tabulated in
Table~\ref{Table3}. The agreement between the analytic and numerical
results is fairly good (at worst the discrepancies amount to $\sim 6$\%)
suggesting that arguments based on Eqs.~(\ref{Taus}), which imply an
expansion in both $\alpha$ and the density around the saturation point
of $N=Z$ matter, are not only insightful but also quantitatively
accurate. The goodness of the parabolic approximation for the binding
energy of asymmetric matter seems to be confirmed also in other
frameworks such as microscopic many-body calculations
\cite{Zuo:2001bd,Heiselberg:1999mq} or model analyses of the symmetry
energy coefficient in nucleus-nucleus collisions \cite{Samaddar:2008gj}.

%%%%%%%%%%%%%%%%%%%%%%%%%%%%%%%%%%%%%%%%%%%%%%%%%%%%%%%%%%%%%%%%%%%%%%
%%%%%%%%%%%%%%%%%%%%%%%%%%%%%%%%%%%%%%%%%%%%%%%%%%%%%%%%%%%%%%%%%%%%%%

We finish this section by revisiting a topic recently addressed in the
literature: {\sl why is Tin so soft?}  \cite{Garg:2006vc,Li:2007bp,
Piekarewicz:2007us,Sagawa:2007sp,Avdeenkov:2008bi}. Namely, GMR
energies of even-$A$ isotopes of Tin from $A=112$ to $A=124$ measured
in a recent experiment \cite{Garg:2006vc,Li:2007bp} are significantly
lower than the values predicted with accurately calibrated, otherwise
successful, mean field models. Note that the same models
satisfactorily predict the GMR excitation energy of ${}^{90}$Zr and
${}^{208}$Pb. Therefore we pose the following question: can the Hybrid
model succeed where the other two (FSUGold and NL3) have failed?
Recall that the Hybrid model was built with the explicit purpose of
having a {\sl ``low''} incompressibility coefficient of
$K_{0}\!\approx\!230$~MeV and a {\sl ``large''} (and negative)
asymmetric term of $K_{\tau}\!=-\!532$~MeV (see Table~\ref{Table3}),
unlike FSUGold where both $K_{0}$ and $|K_{\tau}|$ are low, and
unlike NL3 where both $K_{0}$ and $|K_{\tau}|$ are high. Thus, as in
Ref.~\cite{Piekarewicz:2007us}, the distribution of isoscalar monopole
strength for the even-$A$ Tin isotopes---from ${}^{112}$Sn up to
${}^{124}$Sn---was computed in a relativistic random-phase
approximation (RPA). Details of the method may be found in
Ref.~\cite{Piekarewicz:2001nm}.

%%%%%%%%%%%%%%%%%%%%%%%%%%%%%%%%%%%%%%%%%%%%%%%%%%%%%%%%%%%%%%%%%%%%%%
%%%%%%%%%%%%%%%%%%%%%%%%%%%%%%%%%%%%%%%%%%%%%%%%%%%%%%%%%%%%%%%%%%%%%%

\begin{figure}[ht]
\vspace{0.50in}
\includegraphics[width=5in,angle=0]{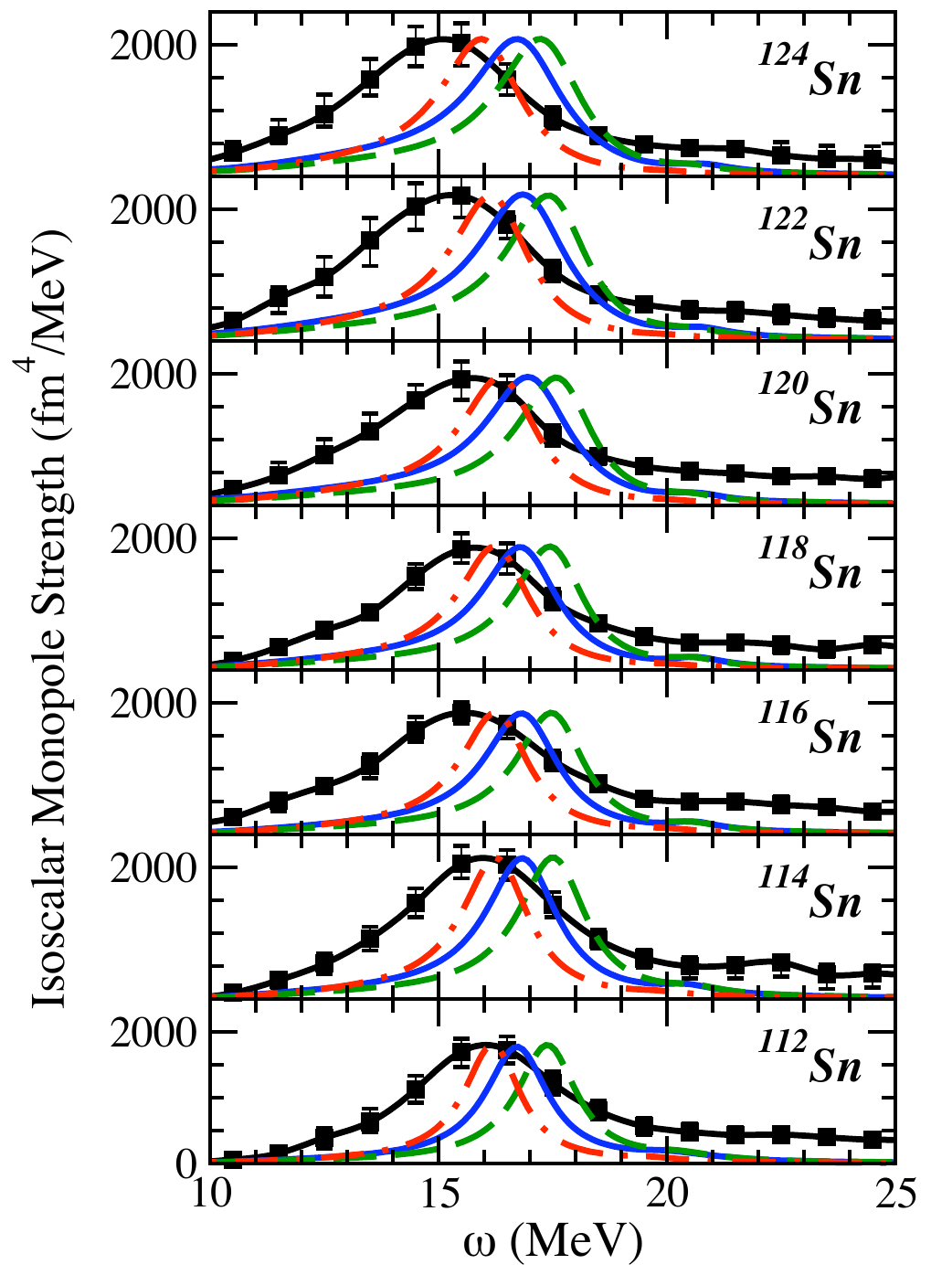}
\caption{(Color online) Comparison between the distribution
          of isoscalar monopole strength in all neutron-even
          ${}^{112}$Sn-${}^{124}$Sn isotopes extracted from
          experiment~\cite{Garg:2006vc,Li:2007bp}
          (black solid squares) and the theoretical
          predictions of the FSUGold (blue solid line), NL3
          (green dashed line), and Hybrid (red dot-dashed line) 
          models.}
\label{Fig7}
\end{figure}

\begin{figure}[ht]
\vspace{0.50in}
\includegraphics[width=5in,angle=0]{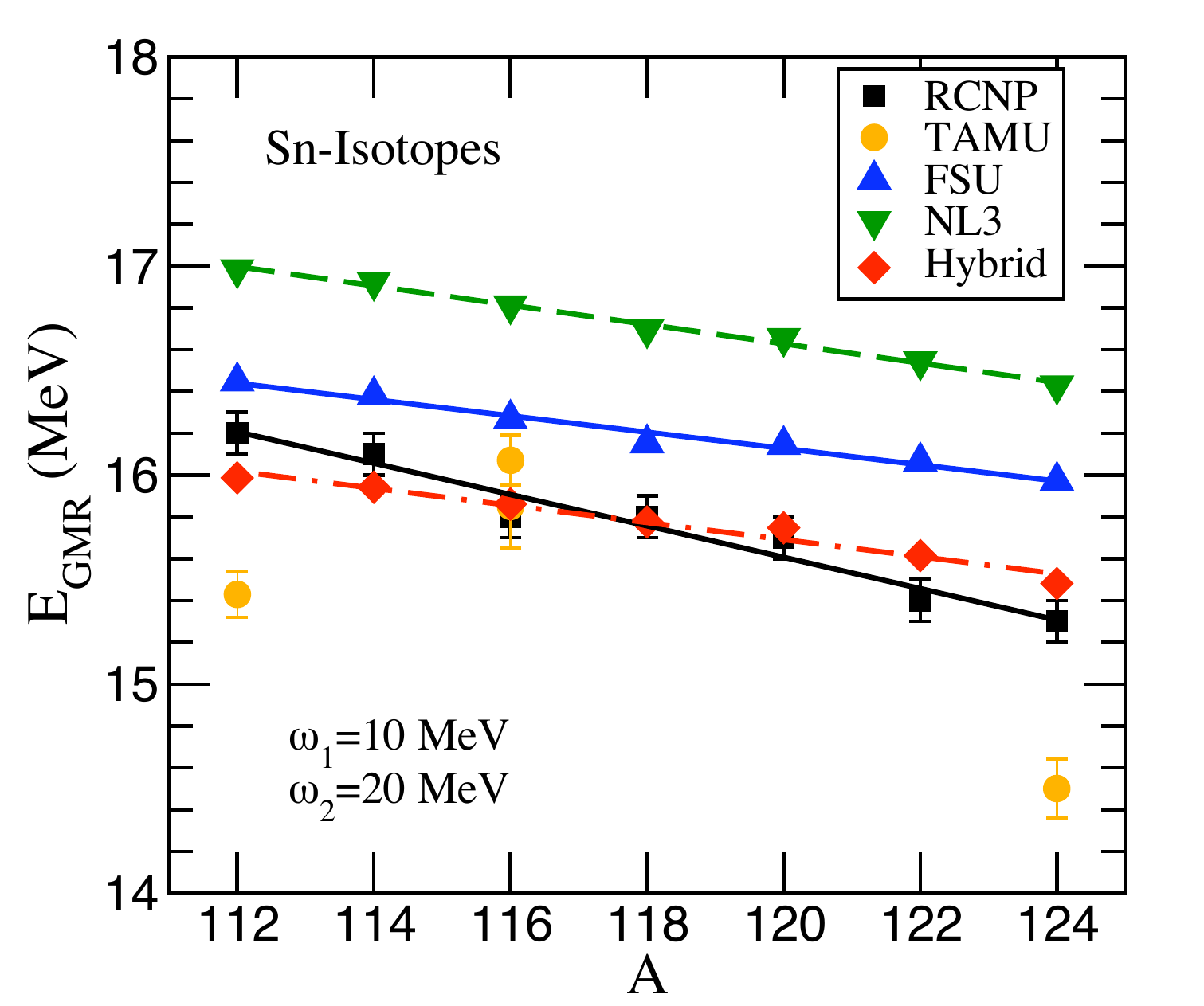}
\caption{(Color online) Comparison between the GMR centroid energies
         ($m_{1}/m_{0}$) in all neutron-even ${}^{112}$Sn-${}^{124}$Sn
         isotopes extracted from experiment~\cite{Garg:2006vc,Li:2007bp}
         (black solid squares) and
         the theoretical predictions of the FSUGold (blue up-triangles),
         NL3 (green down-triangles), and Hybrid (red dot-dashed line)  
         models. Also shown (filled gold circles) are experimental
         results from the Texas A\&M 
         group~\cite{Youngblood:1999,Youngblood:2004fe,Lui:2004wm} for
         the cases of ${}^{112}$Sn, ${}^{116}$Sn, and ${}^{124}$Sn.}
\label{Fig8}
\end{figure}

\begin{figure}[ht]
\includegraphics[width=6in,angle=0]{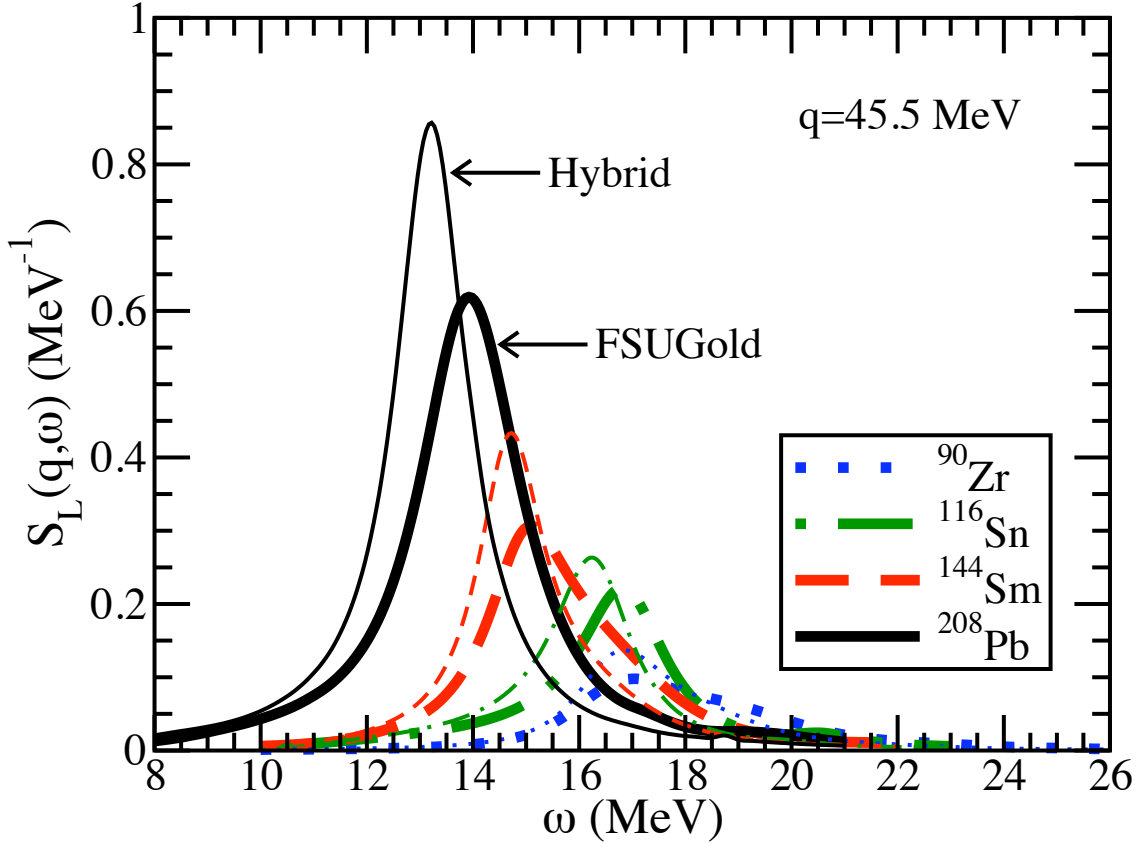}
\caption{(Color online) Distribution of isoscalar monopole strength
         for ${}^{90}$Zr, ${}^{116}$Sn, ${}^{144}$Sm, and ${}^{208}$Pb
         as predicted for the FSUGold (thick lines) and Hybrid (thin 
         lines) models. Experimental centroid energies for these nuclei 
         were reported in Ref.~\cite{Youngblood:1999} and have been 
         tabulated along the theoretical predictions in Table~\ref{Table4}.}
\label{Fig9}
\end{figure}

Figures~\ref{Fig7},~\ref{Fig8}, and~\ref{Fig9} are reminiscent of
those published in Ref.~\cite{Piekarewicz:2007us} but in the present
case results are also included for the Hybrid model. It is clear from
Fig.~\ref{Fig7} that the experimental distribution of strength in the
Tin isotopes is best reproduced by the Hybrid model.  Note that RPA
distributions of strength fail to account for the full --- {\sl
escape-plus-spreading} --- width of the resonance.  Whereas the RPA
calculation accounts properly for the escape width ({\it i.e.}, the
coupling to the continuum is treated exactly) it fails to account for
its spreading component as this one is related to configurations
significantly more complex than those included in the RPA approach.

%%%%%%%%%%%%%%%%%%%%%%%%%%%%%%%%%%%%%%%%%%%%%%%%%%%%%%%%%%%%%%%%%%%%%%
%%%%%%%%%%%%%%%%%%%%%%%%%%%%%%%%%%%%%%%%%%%%%%%%%%%%%%%%%%%%%%%%%%%%%%

On the other hand, the RPA approach is sophisticated enough to 
reproduce the experimental centroid energy of the resonance. As 
it was done experimentally, the centroid energy was computed from 
the ratio of the $m_{1}$ to the $m_{0}$ moment. That is,
%%%
\begin{equation}
 E_{\rm GMR} = \frac{m_{1}}{m_{0}} = \frac{
 \int_{\omega_{1}}^{\omega_{2}} \omega S_{\rm L}(q,\omega) d\omega}
{\int_{\omega_{1}}^{\omega_{2}}        S_{\rm L}(q,\omega) d\omega}\;, 
\end{equation}
%%% 
where $S_{\rm L}(q,\omega)$ is the distribution of strength. The
integration limits have been fixed at $\omega_{1}\!=\!10$~MeV and
$\omega_{2}\!=\!20$~MeV, respectively, and the integrals have been
evaluated at the small momentum transfer of $q\!\sim\!0.23~{\rm
fm}^{-1}$ (or $q\!\sim\!45~{\rm MeV}$)~\cite{Piekarewicz:2007us}. The
theoretical predictions for $E_{\rm GMR}$ in the Tin isotopes are
displayed in Fig.~\ref{Fig8} in comparison with the experimental data
from RCNP~\cite{Garg:2006vc, Li:2007bp} and
TAMU~\cite{Youngblood:1999,Youngblood:2004fe,Lui:2004wm}.

%%%%%%%%%%%%%%%%%%%%%%%%%%%%%%%%%%%%%%%%%%%%%%%%%%%%%%%%%%%%%%%%%%%%%%
%%%%%%%%%%%%%%%%%%%%%%%%%%%%%%%%%%%%%%%%%%%%%%%%%%%%%%%%%%%%%%%%%%%%%%

Although the FSUGold and Hybrid models share the same value of the
incompressibility coefficient in symmetric nuclear matter, the Hybrid
model provides a softer distribution of strength because of its 
largest (negative) asymmetric term $K_{\tau}$ (see Table~\ref{Table3}).
Ultimately, this result hinges on the fact that the Hybrid model has,
as NL3, a significantly stiffer symmetry energy. All in all, the
agreement between the Hybrid model and experiment is adequate,
although the model --- indeed {\sl all models} --- could benefit from 
a steeper slope in the change of the centroid energy $E_{\rm GMR}$ with
mass number $A$.
%%%%%%% ADDED IN RESPONSE TO THE REFEREE'S COMMENTS %%%%%%%
To test the robustness of our results we have used an improved version
of the Hybrid model that was obtained through a slight adjustment of
the scalar mass $m_{s}$ ($508.194\rightarrow 494$~MeV) and the
corresponding coupling constant $g_{\rm s}^2$ ($106.2575\rightarrow
100.4048$, which yields the same $g_{\rm s}^2/m_{s}^2$ value and
ensures that all of the properties of the EOS of infinite nuclear
matter remain unaltered). This mild adjustment yields better
ground-state masses for a few selected nuclei (${}^{40}$Ca,
${}^{90}$Zr, and ${}^{208}$Pb), albeit at the expense of slightly
worsen charge radii. Yet the GMR energies for the Sn-isotopes get
softened by at most 1.5\%. This confirms one of the main results of
this work, namely, that even a model with a soft $K_{0}$ (such as
FSUGold) and a stiff $K_{\tau}$ (such as NL3) is unable to fully
account for the rapid softening of the experimental GMR energies in
the Tin isotopes.
%%%%%%%%%%%%%%%%%%%%%%%%%%%%%%%%%%%%%%%%%%%%%%%%%%%%

%%%%%%%%%%%%%%%%%%%%%%%%%%%%%%%%%%%%%%%%%%%%%%%%%%%%%%%%%%%%%%%%%%%%%%
%%%%%%%%%%%%%%%%%%%%%%%%%%%%%%%%%%%%%%%%%%%%%%%%%%%%%%%%%%%%%%%%%%%%%%

%%%%%%%%%%%%%%%%%%%%%%%%%%%%%%%%%%%%%%%%%%%%%%%%%%%%%%%%%%%%%%%%%%%%%%
%%%%%%%%%%%%%%%%%%%%%%%%%%%%%%%%%%%%%%%%%%%%%%%%%%%%%%%%%%%%%%%%%%%%%%

Therefore, where does theory stand with respect to experiment?
Unquestionably, a Hybrid model having a soft incompressibility
coefficient but a stiff symmetry energy leads to a significant
improvement when compared with the experiment on the Tin
isotopes~\cite{Garg:2006vc,Li:2007bp}. Unfortunately, the Hybrid model
does not fare as well against other observables (see
Table~\ref{Table4}).  First, the Hybrid model predicts a GMR centroid
energy in ${}^{208}$Pb of $E_{\rm GMR}\!=\!13.27$~MeV (or
13.16 MeV if we use the Hybrid model with $m_{s}=494$~MeV),
significantly lower than the experimental value of
$14.17\!\pm\!0.28$~MeV~\cite{Youngblood:1999}; in contrast, the
FSUGold model gives a value of $E_{\rm
GMR}\!=\!14.04$~MeV~\cite{Todd-Rutel:2005fa} that is consistent with
experiment.  Note that similar trends have recently been reported by
Avdeenkov and collaborators~\cite{Avdeenkov:2008bi}.  To appreciate
the significant softening of the Hybrid model relative to FSUGold, the
distribution of isoscalar monopole strength for ${}^{90}$Zr,
${}^{116}$Sn, ${}^{144}$Sm, and ${}^{208}$Pb is displayed in
Fig.~\ref{Fig9}. Second, as we have argued earlier, a large negative
asymmetry term $K_{\tau}$ requires a large positive value of the
asymmetry pressure $L$.  However, models with a stiff symmetry energy
appear to be in conflict with model-independent predictions for the
equation of state of pure neutron matter at low
densities~\cite{Schwenk:2005ka,Horowitz:2005nd,Kowalski:2007,
Piekarewicz:2007dx}. Finally, note that a stiff symmetry energy at
densities below saturation also seems to be disfavored by heavy-ion
data~\cite{Li:2005jy,Li:2008gp,Shetty:2005qp,Shetty:2007zg}, although
in this case the model dependence may be more significant and the
results are not always without some
controversy~\cite{Ono:2005vv,Shetty:2006hh}.

%%%%%%%%%%%%%%%%%%%%%%%%%%%%%%%%%%%%%%%%%%%%%%%%%%%%%%%%%%%%%%%%%
\begin{table}
\begin{tabular}{|c||c|c|c|c|c|c|}
 \hline
 Nucleus & $\alpha$ & $\omega_{1}-\omega_{2}$ & Experiment & FSU & NL3 & Hybrid\\
 \hline
 ${}^{90}$Zr  & $0.111$ & $10-26$ & $17.89\pm0.20$ & $17.98$ & $18.62$ & $17.47$ \\
 ${}^{116}$Sn & $0.138$ & $10-23$ & $16.07\pm0.12$ & $16.58$ & $17.10$ & $16.02$ \\
 ${}^{144}$Sm & $0.139$ & $10-22$ & $15.39\pm0.28$ & $15.64$ & $16.14$ & $15.07$ \\
 ${}^{208}$Pb & $0.212$ & $8-21$ & $14.17\pm0.28$ & $14.04$ & $14.32$ & $13.27$ \\
\hline
\end{tabular}
\caption{Giant-Monopole-Resonance centroid energies ($E_{\rm
GMR}\!\equiv\!m_{1}^{}/m_{0}^{}$ in MeV) obtained from the 
distribution of monopole strength integrated over the 
$\omega_{1}$--$\omega_{2}$ range for those nuclei studied in
Ref.~\cite{Youngblood:1999}; $\alpha$ represents their respective neutron-proton asymmetry.}
\label{Table4}
\end{table}
%%%%%%%%%%%%%%%%%%%%%%%%%%%%%%%%%%%%%%%%%%%%%%%%%%%%%%%%%%%%%%%%%

\section{Conclusions}
\label{Conclusions} 

The saturation properties of neutron-rich matter were studied as a
function of the neutron-proton asymmetry within the framework of
relativistic mean-field models. We observed that infinite nuclear
matter continues to saturate up to values of the neutron-proton
asymmetry of the order of
$\alpha\!\equiv\!(N-Z)/A\!\lesssim\!0.75$. Moreover, it was found
quite generally that as infinite nuclear matter departs from the
symmetric $(N\!=\!Z)$ limit, the saturation density lowers, the
binding energy weakens, and the nuclear incompressibility softens.

The manuscript was organized around three main themes: (a) the use of
accurately calibrated relativistic mean-field models to extract the
saturation properties of neutron-rich matter directly from numerical
computations and the comparison of these numerical results against
approximate analytic approaches; (b) the use of the same models to
compute the distribution of isoscalar-monopole strength in various
nuclei; and (c) the comparison of these theoretical predictions
against the experimentally measured GMR energies.

To make contact between the equation of state of bulk neutron-rich
matter and GMR energies on finite nuclei, the incompressibility of
infinite neutron-rich matter was parametrized in terms of two bulk
parameters, namely, $K_{0}$ and $K_{\tau}$, with the former being the
incompressibility coefficient of \textsl{symmetric} matter and the
latter parametrizing the (small) deviations from the symmetric limit
[see Eq.~(\ref{KTau})]. Note that never in the manuscript we relied on
semi-empirical (\textsl{liquid-drop-like}) formulas to extract
properties of infinite-matter from extrapolating finite-nuclei results
to the $A\!\rightarrow\infty$ limit. In this manner we followed
the time-honored tradition initiated by Blaizot and 
collaborators~\cite{Blaizot:1976,Blaizot:1995} of demanding that 
the values of both $K_{0}$ and $K_{\tau}$ be those extracted from a 
consistent theoretical model that successfully reproduces the 
experimental GMR energies of a variety of nuclei.

As part of the first theme, the evolution with neutron-proton
asymmetry of the saturation density, binding energy per nucleon, and
incompressibility coefficient were extracted from a fit to the
numerically generated equation of state. Once these properties were
extracted, their dependence on the neutron-proton asymmetry $\alpha$ 
was captured through a simple parametrization in powers of $\alpha^{2}$ 
with no reliance on the parabolic approximation of Eq.~(\ref{EOS}) nor 
on an expansion involving bulk-model parameters,
as in Eq.~(\ref{EvsX}). Having completed this numerical procedure,
we explored the possibility of reproducing the exact
numerical results from analytic expansions based on a few bulk
parameters of the equation of state determined at normal
nuclear-matter saturation density [see Eqs.~(\ref{Taus})].
For all three bulk properties --- the saturation density, the binding
energy per nucleon, and the incompressibility coefficient --- the
analytic values were in close agreement with those computed
numerically. This seems to be a robust result as it holds for all
three (FSUGold, NL3, and Hybrid) models; see Table~\ref{Table3}.
Thus, we concluded that the analytic expressions are not only
insightful but also quantitatively accurate.  Particularly interesting
was the case of the incompressibility coefficient $K_{\tau}$ that is
given as the sum of three potentially \textsl{``large and
cancelling''} contributions. However, we found that one of these three
terms --- the slope of the symmetry energy $L$ --- dominates
$K_{\tau}$ [see Eq.~(\ref{KTau})], thereby making sensitive
cancellations unlikely. This result revealed an interesting
correlation between $K_{\tau}$ and $L$ that may be further explored by
the upcoming Parity Radius Experiment (PREx) at the Jefferson
laboratory.  PREx promises to measure the neutron radius of $^{208}$Pb
accurately and model independently via parity-violating electron
scattering~\cite{Horowitz:1999fk, Michaels:2005}. PREx will provide a
unique experimental constraint on the density dependence of the
symmetry energy due its strong correlation to the neutron radius (or
neutron skin thickness) of heavy nuclei~\cite{Brown:2000}.

To test the predictions of these three models contact had to be made
with available experimental data on finite nuclei. Thus, the
distribution of isoscalar monopole strength was computed for a variety
of nuclei in a consistent RPA approach~\cite{Piekarewicz:2001nm}. In
particular, the main motivation behind introducing the \textsl{Hybrid}
model was due to the inability of both FSUGold and NL3 to reproduce
the recently-measured GMR energies 
along the isotopic chain in
Tin~\cite{Garg:2006vc,Li:2007bp}. By adopting a relatively small value
for the incompressibility coefficient in symmetric matter
($K_{0}=\!230\!$~MeV) together with a fairly large negative value for
its leading deviation from the symmetric limit
($K_{\tau}\!\approx\!-530$~MeV), we were constructing a Hybrid model
with a significantly softer incompressibility coefficient for
neutron-rich matter.  Such a softening indeed produced a significant
improvement \textsl{vis-\`a-vis} the experimental data on the
Sn-isotopes; see Fig.~\ref{Fig8}. Whereas FSUGold and NL3 overestimate
the centroid energy in ${}^{124}$Sn by about $0.7$ and $1.0$ MeV,
respectively, the Hybrid model falls within $0.1$ MeV of the
experimental data. Indeed, the predictions of the Hybrid model fall
within $0.1$~MeV of the experimental data for the \textsl{full
isotopic chain} if one takes account of the uncertainties in
the data. However, although the improvement in the case of the
Sn-isotopes is significant and unquestionable, an important problem
remains: the Hybrid model \textsl{underestimates} the GMR centroid
energy in ${}^{208}$Pb --- the heaviest doubly-magic nucleus --- by
almost 1 MeV. This suggests that the rapid softening with neutron
excess predicted by the Hybrid model may be unrealistic.

Thus, where does theory stand with respect to experiment? One
possibility, given that FSUGold reproduces the centroid energy in both
${}^{90}$Zr (with $\alpha\!=\!0.11$) and ${}^{208}$Pb (with
$\alpha\!=\!0.21$), is that its predictions for $K_{0}$ and $K_{\tau}$
are reliable but that its failure to reproduce the GMR energies in Tin
is due to missing physics unrelated to the incompressibility of
neutron-rich matter. We feel inclined to favor this possibility
because of two main reasons. First, the missing physics may be to some
extent related to the open-shell structure of the Tin-isotopes, a
property that makes pairing correlations essential and endows Tin with
its superfluid character. Support in favor of this scenario has been
recently presented in Ref.~\cite{Li:2008hx} where a surface pairing
force was used to bring theory much closer to experiment, at least
from ${}^{112}$Sn to ${}^{120}$Sn. Second, the large and negative
value suggested from the experimental extraction of $K_{\tau}$ may be
at odds with theoretical constraints deduced from the behavior of
dilute Fermi gases that seem to suggest a moderate value for the
pressure of pure neutron matter at saturation
density~\cite{Piekarewicz:2007dx}. (Note that the pressure of pure
neutron matter, or equivalently the slope of the symmetry energy $L$,
largely determines the behavior of $K_{\tau}$.)  This suggests that
the value of $K_{\tau}\!=\!-550\!\pm\!100$~MeV inferred from
experiment~\cite{Garg:2006vc,Li:2007bp} may suffer from the same
ambiguities already encountered in earlier attempts to extract the
incompressibility coefficient of \textsl{infinite matter} from
finite-nuclei extrapolations. Yet the final resolution on
\textsl{``why is Tin so soft?''} awaits further theoretical insights.

\vfill\eject

\begin{acknowledgments}
 The authors are grateful to Professors M. Di Toro and H. Sagawa for
 useful discussions. This work was supported in part by grants from
 the U.S. Department of Energy DE-FD05-92ER40750, FIS2008-01661 from
 MEC (Spain) and FEDER, and 2005SGR-00343 from Generalitat de
 Catalunya, and by the Spanish Consolider-Ingenio 2010 Programme CPAN
 CSD2007-00042.
\end{acknowledgments}

%%%%%%%%%%%%%%%%%%%%%%%% START THE APPENDICES %%%%%%%%%%%%%%%%%%%
\appendix

\section{Nomenclature and Terminology}
\label{Nomenclature} 

In this section we address what we perceive as a confusing state of
affairs in regard to the nomenclature used to characterize the
symmetry energy. First, we note that no uniform terminology exists
even to denote the neutron-proton asymmetry coefficient. Indeed, the
symbols $I$~\cite{Baran:2004ih},
$\alpha$~\cite{Li:2008gp,Furnstahl:2001un,Chen:2007ih}, 
$\beta$~\cite{Zuo:2001bd}, and
$\delta$~\cite{Li:2005jy,Blaizot:1980tw,Sagawa:2007sp} are all used in
the literature to denote the neutron-proton asymmetry coefficient
$(N-Z)/A$ of asymmetric nuclear matter. Second, and perhaps even more
confusing, is the myriad of different symbols used to refer to the
same bulk properties. For example, all of the following expressions 
may be found in the literature~\cite{Baran:2004ih,Furnstahl:2001un}:
%%%
\begin{subequations}
 \begin{align}
  {\cal S}(\rho) 
  &= J + Lx + \frac{1}{2}K_{\rm sym}x^{2}+\ldots \\
  &= a_{4}  
  + \frac{L}{3}
    \left(\frac{\rho-\rho_{{}_{0}}}{\rho_{{}_{0}}}\right) 
  + \frac{K_{\rm sym}}{18}
    \left(\frac{\rho-\rho_{{}_{0}}}{\rho_{{}_{0}}}\right)^{2}
  + \ldots \\
  &= a_{4}+\frac{P_{0}}{\rho_{{}_{0}}^{2}}(\rho-\rho_{{}_{0}})+
    \frac{\Delta K_{0}}{18\rho_{{}_{0}}^{2}}(\rho-\rho_{{}_{0}})^{2}
  + \ldots\;,
 \end{align}
 \label{SymmEMess}
\end {subequations} 
%%% 
where $x\!=\!(\rho\!-\!\rho_{{}_{0}})/3\rho_{{}_{0}}$. Moreover, another
plausible expansion of the symmetry energy may be around the 
equilibrium {\sl Fermi momentum}~\cite{Blaizot:1980tw}. That is,
%%%
\begin{equation}
  {\cal S}(\rho) 
   = \tilde{J} + \tilde{L}y + \frac{1}{2}\widetilde{K}_{\rm sym}y^{2}
   +\ldots\;, \\
 \label{SymmEkFermi}
\end {equation} 
%%% 
where the deviation from the equilibrium Fermi momentum has been
parametrized in terms of the dimensionless parameter $y$ defined 
as
%%%
\begin{equation}
 y \equiv \frac{(k_{\rm F}-k_{\rm F}^{0})}{k_{\rm F}^{0}} \;.
 \label{YParam}
\end {equation} 
%%% 
Recall that the Fermi momentum and the baryon density are related by
the following expression:
%%%
\begin{equation}
 \rho = \frac{2k_{\rm F}^{3}}{3\pi^{2}} \;.
 \label{RhovskFermi}
\end {equation} 
%%% 
One potential confusion between the two different expansions of the
symmetry energy (in terms of either $x$ or $y$) is that in the
presence of a linear term (such as $L$) the coefficients are in
general not equal. Indeed, the various bulk coefficients are 
related as follows:
%%%
\begin{subequations}
 \begin{align}
   & \tilde{J}=J \;, \\
   & \tilde{L}=L \;, \\
   & \widetilde{K}_{\rm sym}= K_{\rm sym}+2L \;.
 \end{align}
 \label{RhovskFermi2}
\end {subequations} 
%%% 
That is, at order $y^{2}$ and higher, the expansion coefficients 
in terms of the Fermi momentum $y$ differ from the corresponding
ones used in an expansion in terms of the density $x$.

It is also common practice to express the {\sl finite nucleus}
incompressibility coefficient ($K_{A}$) by means of a liquid-drop-like
mass formula~\cite{Blaizot:1980tw,Blaizot:1995,Chossy:1997,Patra:2001mt,
Li:2007bp,Sagawa:2007sp,Chen:2007ih}, which highlights the physical
meaning of the various contributions to $K_{A}$. That is,
%%%
\begin{equation}
 K_{A} = K_{\rm vol} + K_{\rm surf}A^{-1/3} 
       + K_{\tau}\left(\frac{N-Z}{A}\right)^{2}
       + K_{\rm Coul}\frac{Z^{2}}{A^{4/3}}
       + \ldots\;.
 \label{KA}
\end {equation} 
%%% 
In some works, the coefficient $K_{\tau}$ is denoted by $K_{\rm asy}$
\cite{Li:2005jy,Li:2008gp,Chen:2007ih} or $K_{\rm vs}$
\cite{Chossy:1997,Patra:2001mt}. To add to the confusion in notation, in
the original contributions by Blaizot and
collaborators~\cite{Blaizot:1976,Blaizot:1980tw,Blaizot:1995} the term
$K_{\rm sym}$ was used instead of $K_{\tau}$ in Eq.~(\ref{KA}). It
appears that at present $K_{\rm sym}$ has been ``universally'' adopted
to refer to the curvature of the symmetry energy at saturation
density, as in Eq.~(\ref{SymmE}).

In summary, we adopt the following convention in the present
manuscript --- and hopefully in all future works. The
energy-per-particle of asymmetric nuclear matter is denoted 
as follows: 
%%%
\begin{align}
 {\cal E}(\rho,\alpha) &= {\cal E}_{\rm SNM}(\rho)+ 
                          {\cal S}(\rho)\alpha^{2}+
                          {\cal O}(\alpha^{4})\nonumber\\
                       &= \Big(\varepsilon_{{}_{0}}+
                          \frac{1}{2}K_{0}x^{2}+
                          \frac{1}{6}Q_{0}x^{3}+\ldots\Big)\nonumber\\
                       &+ \Big(J+Lx+\frac{1}{2}K_{\rm sym}x^{2}+
                          \frac{1}{6}Q_{\rm sym}x^{3}+\ldots\Big)\alpha^{2}
                          + {\cal O}(\alpha^{4})\;,
 \label{Expansion}
\end{align} 
%%%
where the dimensionless parameters $x$ and $\alpha$ characterize the
deviations from saturation density and from the symmetric limit,
respectively. That is, 
%%%
\begin{align}
  & x \equiv \frac{(\rho-\rho_{{}_{0}})}{3\rho_{{}_{0}}} \;, \\
  & \alpha \equiv \frac{(N-Z)}{A} \;.
 \label{XandAlphaParams}
\end{align} 
%%%
Finally, the quantities $\rho_{\tau}$, $\varepsilon_{\tau}$, and
$K_{\tau}$, have been introduced to denote, respectively, the
evolution with the neutron-proton asymmetry of the saturation density,
the energy per particle, and the incompressibility coefficient of
\textsl{infinite} neutron-rich matter. That is,
%%%
\begin{align}
  & \rho_{{}_{0}}(\alpha)=\rho_{{}_{0}}+\rho_{\tau}\alpha^{2}
               +{\cal O}(\alpha^{4})\;, \\
  & \varepsilon_{{}_{0}}(\alpha)=\varepsilon_{{}_{0}}
               +\varepsilon_{\tau}\alpha^{2}
               +{\cal O}(\alpha^{4})\;, \\
  & K_{0}(\alpha)=K_{0}+K_{\tau}\alpha^{2}+{\cal O}(\alpha^{4})\;.
 \label{TauParams}
\end{align} 
%%%

\vfill\eject
\bibliography{/Users/jorge/Tex/Papers/ReferencesJP}

\end{document}